# Assessing salt precipitation and weak acid interaction in subsurface $CO_2$ injection: Potential 50% strength decline in near-wellbore reservoir sandstones

Mohammad Nooraiepour[1], Krzysztof Polański[2], Mohammad Masoudi[1], Szymon Kuczyński[2], Hannelore Derluyn[3], Liebert Parreiras Nogueira[4], Bahman Bohloli[5], Stanislaw Nagy[2] and Helge Hellevang[1]

[1] Department of Geosciences, University of Oslo, P.O. Box 1047 Blindern, 0316 Oslo, Norway
[2] AGH University of Kraków, Mickiewicza 30 Av, 30-059 Kraków, Poland
[3] Universite de Pau et des Pays de l'Adour, E2S UPPA, CNRS, LFCR, Pau, France
[4] Oral Research Laboratory, Institute of Clinical Dentistry, University of Oslo, Oslo 0317, Norway
[5] Norwegian Geotechnical Institute (NGI), Sognsveien 72, Oslo, 0855, Norway

* Corresponding author: mohammad.nooraiepour@geo.uio.no

## Abstract
Predictive modeling of $CO_2$ storage sites requires a detailed understanding of physico-chemical processes and potential challenges for scale-up. Dramatic injectivity decline may occur due to salt precipitation pore clogging in high-salinity reservoirs, even over a short time frame. This study aims to elucidate the adverse impact of $CO_2$-induced salt crystallization in porous media on the geomechanical properties of near-wellbore reservoir sandstones. As the impact of salt precipitation cannot be isolated from the precursor effects of interaction with $CO_2$ and carbonic acid, we initiated our study by comprehensive review of $CO_2$ chemo-mechanical interactions with sandstones. We conducted laboratory geochemical $CO_2$-brine-rock interactions at elevated pressures and temperatures on two sets of porous sandstone with contrasting petrophysical qualities. Two paths were followed: treatment with (a) $CO_2$-acidified brine at 10 MPa fluid pressure and 60 °C for 7 days, and a second subset continuation with (b) supercritical injection until complete dry-out and salt precipitation. Afterward, the core samples were tested in a triaxial apparatus at varying stresses and temperatures. The elastic moduli of intact, $CO_2$-reacted, and salt-damaged sandstones were juxtaposed to elucidate the extent of crystallization damages. The salt-affected specimens showed a maximum of 50% reduction in Young's and shear moduli and twice an increase in Poisson's ratio compared to intact condition. The deterioration was notably higher for the tighter rocks with higher initial stiffness.

Keywords: Review; Salt precipitation; Geochemical reaction; Fluid-Rock interaction; Geomechanical properties; Geological carbon storage

## 1. Introduction
Achieving a significant and timely reduction in global greenhouse gas emissions is a major challenge for modern human civilization. There is growing consensus that an essential part of the solution to mitigate climate change is the deployment of long-term subsurface storage of carbon dioxide ($CO_2$) in geological formations (IPCC, 2023), a technological chain of processes referred to as Carbon Capture and Storage (CCS), Geological $CO_2$ Storage (GCS), or $CO_2$ sequestration. Among the various opportunities for atmospheric $CO_2$ emission reductions, such as a blend of more renewable energy, improved energy efficiency, CCS contribution is anticipated to be around 10-15% of total cumulative emissions reductions through 2050, which is approximately 120 Gigatons (Gt) (IEA, 2016; Ringrose et al., 2021). The IPCC AR6 Synthesis Report refers to CCS technology as a critical $CO_2$ mitigation option. In 1.5°C Special Report, three of the four pathways involve significant use of CCS (350 to 1200 gigatons stored by 2100), mandating radical changes in human behavior in the last pathway without CCS deployment. The required subsurface storage rate need to be 6–7,000 million tons annually in 2050 (IEA, 2015; Ringrose and Meckel, 2019). It has been argued that GCS in saline aquifers enables delivering the long-term expectations for CCS and realizing the 2-degree scenario (2DS) goals (Ringrose and Meckel, 2019), while GCS in petroleum reservoirs and mafic/ultramafic rocks (e.g., basalt) will play an important role in some geographic locations.

Given the expected enormous annual injection of $CO_2$ in saline aquifers with varying salinity and petrophysical properties, it is critical to delineate dominant physicochemical processes that occur during saline aquifer storage. Injected $CO_2$ in geological formations disturbs the existing equilibrium between formation brine and rock-forming minerals, acidifies the resident brine, triggers (geo)chemical reactions, and accelerates fluid-rock interactions in porous media, which may alter the mechanical and hydraulic properties of the reservoir and caprock layers (Liu et al., 2012; Masoudi et al., 2024; Nooraiepour et al., 2022; Pham et al., 2011; Rohmer et al., 2016; Vafaie et al., 2023). Out of potential thermo-hydro-mechanical-chemical (THMC) coupling during GCS, we focus on chemically-induced alterations of rocks. In particular, changes in geomechanical properties of reservoir rocks in hypersaline aquifers in the regions near injection wells. The distinction between the consequences of $CO_2$-induced salt precipitation in porous media that comes into effect after collective impacts of interactions with pure $CO_2$ and $CO_2$-acidified brine is challenging to establish reliably. Our introduction thus initiates with a comprehensive literature review, followed by a parallel exploration in our experiments.

Table 1 (Refer to Appendix) presents an overview of research on chemo-mechanical interactions of $CO_2$ (dry, wet, and water/brine-acidified) only with sandstone rock type. The literature indicates that the geomechanical properties of $CO_2$-reacted sandstones experience deterioration. This includes elastic parameters (Young's modulus, bulk modulus, shear modulus, and Poisson's ratio), strength, failure and deformation responses, creep properties, stress-strain behavior, and rock physics properties. This degradation is particularly noticeable when geochemical reactions (involving fluid-rock interactions) occur under elevated temperature-pressure conditions, such as those found at reservoir conditions, and extend over prolonged experimental periods. However, the above-mentioned consequences and their severity are not unanimous, particularly when carbonate-lean sandstones (as clastic grains or intergranular cement) are evaluated (refer to Table 1). The decline in mechanical strength, often rooted in the debilitation of grain bonding, results in the geomechanical weakening of reservoir rocks. The presence of impurities such as SOx, NOx, in addition to supercritical $CO_2$ ($ScCO_2$) has been shown to result in enhanced weakening in rock mechanical properties (Erickson et al., 2015). Subsurface $CO_2$ injection also changes the stress regime in the storage reservoirs and overlying caprocks, particularly in the near-wellbore regions (Rahman et al., 2022; Rutqvist et al., 2008; Vilarrasa et al., 2014). Deteriorated strength and stiffness parameters (physical and mechanical properties) come into play when evaluating safe and secure GCS (Nooraiepour et al., 2017; Song and Zhang, 2013; Watts, 1987).

Type, extent, and kinetics of chemical reactions (or fluid-rock interactions) may vary across zones with different distances from the wellbores as a function of the volumetric fraction of aqueous phases (brine) and $CO_2$ in porous media, and the mutual concentration of dissolved water and $CO_2$ in each other (Rohmer et al., 2016; Vafaie et al., 2023). In other words, fluid phase saturation in pore structures controls interactions (Kampman et al., 2014; Miri and Hellevang, 2016; Nooraiepour, 2018). In the nearest region to the wellbore, dry $ScCO_2$ occupies most of the reservoir pore volume. The low chemical reactivity of $ScCO_2$ during continuous injection may sweep out the brine pore fluid, dry out the region, and cause precipitation and growth of salt crystals (Masoudi et al., 2021; Miri and Hellevang, 2016; M Nooraiepour et al., 2019). The next region around the injection well represents a zone of water-bearing $ScCO_2$ or saturated/wet $ScCO_2$. In contrast to dry $ScCO_2$, it shows a reactive nature enhancing surface/interface (geo)chemical processes on rock-forming minerals (Lin et al., 2008; Loring et al., 2011; Pearce et al., 2016). This zone is followed by an extended region where two fluids (brine and $ScCO_2$) coexist at varying proportions. In all these three regions, $CO_2$ dissolution in brine leads to carbonic acid generation ($H_2CO_3$), which dissociates into aqueous protons ($H^+$), leading to a pH drop to an acidic range of 3.5–5.0 (Gaus, 2010). It, in turn, induces subsequent surface/interface-coupled mineral dissolution and precipitation (An et al., 2021; Chen et al., 2014; Deng et al., 2022; Fazeli et al., 2019b, 2019a; Guren et al., 2020; Prasianakis et al., 2017).

Acidified pore water can cause changes in the pore geometry of reservoir rocks and bring about mineral dissolution and precipitation. The resulting pore morphology alteration affects the rock porosity and permeability (i.e., porous media hydrodynamics) and, eventually, the mechanical properties and strength of the targeted storage sites (Kim and Santamarina, 2014; Nooraiepour, 2018; Vafaie et al., 2023; Zhang et al., 2016). The volumetric content and spatial distribution of reactive minerals, in addition to the porous structure and fluid flow condition, will determine which rock properties may change. $CO_2$-fluid-rock interactions have been thoroughly studied in the past two decades. For instance, refer to the reviews conducted by Czernichowski-Lauriol et al. (1996), Gaus (2010), Zhang and Bachu (2011), Song and Zhang (2013), Rohmer et al. (2016), and Vafaie et al. (2023). Interaction with $CO_2$-acidified water leads to enhanced chemical reactions in carbonates, whereas the acidity of the injected fluid is expected to have a less adverse impact on anhydrites, clay-rich caprocks, and sandstones.

The composition of intergranular cement in sandstones can play a remarkable role in the rock properties evolution, i.e., silicate or calcite cementation. Carbonate-cemented reservoir sandstones are susceptible to enhancements in dissolution-induced porosity (and permeability) (Espinoza et al., 2018; Fazeli et al., 2019a; Rathnaweera et al., 2017; Wu et al., 2018). In particular, under advective (open flow) conditions hindering buffering effects (Fazeli et al., 2019a, 2019b). Additionally, studying outcrops and natural-analog specimens exposed to $CO_2$ migration over geological time documented a porosity increase due to the dissolution of grain-coating hematite and calcite cement. The dissolution rate of sandstone grains, which often constitute quartz, feldspar, and clay grains are markedly slower than carbonates, particularly calcite (Brantley et al., 2008; Palandri and Kharaka, 2004). Nevertheless, as documented in several laboratory and numerical experiments, low-carbonate sandstones containing clays and feldspars are more reactive with dissolved $CO_2$ over longer times than quartz grains (Hangx and Spiers, 2009; Pham et al., 2011; Tutolo et al., 2015; Zhu et al., 2019).

A $CO_2$-brine-rock interaction that is rapid and relevant for injection-timescales, is the $CO_2$-induced salt precipitation, which is geochemical reaction with potentially significant consequences in the near-wellbore regions is $CO_2$-induced salt precipitation. Injection of large volumes of undersaturated $ScCO_2$ leads to the evaporation of the formation water and an increase in the concentration of the dissolved salts in brine pore fluid. Under the thermodynamic conditions of a given storage reservoir, solute concentration will eventually reach the solubility limit and precipitate out of the aqueous phase, leading to salt precipitation inside the porous reservoir rocks (Falcon-Suarez et al., 2020; Miri and Hellevang, 2016; Mohammad Nooraiepour et al., 2018; Norouzi et al., 2022; Ott et al., 2021; Parvin et al., 2020). This study primarily investigates subflorescence (salt crystallization inside the material), contrasting efflorescence (surface growth).

In construction, rock slope stability, and cultural heritage preservation, salt crystallization in porous materials, often called salt weathering, is considered one of the significant causes of rock decay in nature and man-made materials (Desarnaud et al., 2013; Flatt et al., 2017; Lubelli et al., 2023). It has been shown that salt precipitation in porous media and microcavities (representing confined spaces) has the potential to exert sufficient stress to overcome the tensile strength of most rocks, alter their pore space, induce microcracks, and change mechanical properties (Derluyn et al., 2014; Desarnaud et al., 2016; Shahidzadeh-Bonn et al., 2010). During surface salt weathering, repeated dissolution-recrystallization of salt in rock pore volumes is the root cause of rock damage, fragmentation and disintegration (Derluyn et al., 2019; Godts et al., 2021). However, not much research has been published on the potential adverse impact of $CO_2$-induced salt precipitation on the geomechanical properties of reservoir rocks in the near-wellbore vicinity. There is also literature speculating that mineral precipitation in porous media during GCS operation may instead increase strength (Rosenbauer et al., 2005).

In high-salinity GCS reservoirs (i.e., hypersaline aquifers), dramatic injectivity decline might be expected due to salt precipitation and continued growth resulting in pore clogging, even in the short-term. Therefore, this study aims to address this knowledge need and elucidate the discrepancy in the positive/negative impact of mineral precipitation (particularly salt subflorescence) in porous media on the geomechanical properties of storage reservoirs. We conducted batch-type interaction of sandstone samples with $ScCO_2$-acidified high salinity sodium chloride brine, and then flow-through $ScCO_2$ injection until salt crystallization and dry-out of porous rocks both at elevated temperatures. Mechanical behavior and elastic moduli of unreacted (intact) and two sets of reacted (i. with $ScCO_2$-acidified brine, and ii. those followed until salt precipitation) sandstone cores were then compared using high-pressure high-temperature (HPHT) triaxial geomechanical experiments. Additionally, two sets of distinct petrophysical quality representing diverse pore structures were studied to provide insight into the range of impacts of in-pore salt crystallization and growth.

## 2. Materials and Methods

### 2.1. Sample preparation

Two Berea-class sandstone core samples with distinct petrophysical quality were selected. The porosity ($\varphi$) and permeability (k) for the Boise Buff (BB) sandstones are 25-28% and 1000 mD ($\sigma$= 700-3000 mD), representing excellent reservoir quality index (RQI). The respective petrophysical properties for the Torrey Red (TR) core samples with low to medium RQI are $\varphi$= 13-16% and 1 mD ($\sigma$= 0.2-2 mD). For each reservoir rock type, we acquired ten cylindrical core plugs perpendicular to the bedding with dimensions of approximately 38.1 mm (1 ½”) in diameter and approximately 76.2 mm (3”) in height.

### 2.2. Sample characterization

The whole-rock (bulk) mineralogical composition was identified and quantified using X-ray diffraction (XRD) technique. The composition was verified using an in-house automated mineralogy and petrography mapping module under energy-dispersive x-ray spectroscopy (EDS) integrated into scanning electron microscopy (SEM). The EDS analysis was used for the identification of chemical elements and surface composition mapping. High-resolution SEM imaging via backscattered (BSE) and secondary electrons (SE) was carried out to study grain-pore distribution, surface geometries in addition to salt growth. The detailed XRD and SEM-EDS procedures have been given in Nooraiepour et al. (2017a,b) and Nooraiepour et al. (2021a,b), respectively.

Three-dimensional (3D) tomograms were recorded for intact, $CO_2$-reacted, and salt-damaged sandstones. The x-ray computed tomography (micro-CT) was performed using a Bruker Multiscale SkyScan 2211 at 110 kV, 65 μA, and an exposure time of 400 ms, averaging 5 frames per projection. The entire set of tomograms was taken over a 360° rotation with a rotation step of 0.29° (1242 projections). These settings resulted in a final voxel size of 12 μm. Micro-CT projections were reconstructed using the proprietary NRecon hierarchical 3D reconstruction software, including ring artifact and beam hardening correction. The resulting reconstructions were then visualized and analyzed using the Dragonfly software (v. 2022.2, Comet Technologies Canada Inc.). The virtual slices were pre-processed to remove distorted outer regions and filtered with artifact attenuation (gradient domain fusion), and non-local means denoising filters. The slices were then segmented into distinct phases with visual thresholds (Otsu algorithm) to define seeds for the watershed algorithm based on the grey-scale gradient and grey-scale intensity of each voxel. The outcome was then used for quantifying each phase's spatial distribution and volumes.

### 2.3. Geochemical fluid-rock interaction

In addition to performing geomechanical experiments on intact (unreacted) BB and TR sandstones, the effects of potential geochemical reactions with $CO_2$-acidified brine on the mechanical properties of $CO_2$ storage reservoir rocks were studied. A total of 10 natural core samples were tested from BB and TR reservoir units. Following the testing of 3/4 intact samples, we selected independent BB and TR sets, each comprising 3 core samples, to explore processes that may take place during $CO_2$ sequestration in high-salinity aquifers:

I. Geochemical reaction with $CO_2$-acidified brine (pressurized supercritical $CO_2$ dissolved in high-salinity brine) over time at high-pressure high-temperature (HPHT) condition

II. Continuation of path I with the injection of supercritical $CO_2$ until evaporation of brine saturation, salt crystal precipitation and growth, and eventually dry out of core sample

Reaction path I simulates generic conditions in hypersaline aquifers, whereas path II focuses on probable scenarios near the $CO_2$ injection well. Four of the ten core samples prepared from BB and TR sandstones were tested at intact (unreacted) condition and six were tested after geochemical reactions (three samples per reaction path).

We used an HPHT core flooding system (AFS 200, Core Laboratories) to carry out brine-rock chemical interactions with pressurized $CO_2$ under reservoir conditions. The fluid flow apparatus is equipped with a forced convection benchtop oven (Despatch LBB series), which combines horizontal and vertical airflow and provides temperature uniformity within the air bath. The fluid injection system comprises multiple syringe pumps (Teledyne Isco) supplemented with a backpressure regulator and overburden pressure pump. Two stainless steel fluid accumulators (transfer vessel) and a Hassler-type core holder within the oven enabled us to equilibrate and inject brine and $ScCO_2$ phases during paths I and II geochemical interactions. For further details of the experimental setup, refer to Moghadam et al. (2019) and Nooraiepour et al. (2018).

An aqueous solution of 3.4 M (200 g/L) sodium chloride was prepared by adding NaCl (ACS reagent grade, EMSURE) to deionized water (Milli-Q water). A bottle provided grade 5.2 scientific carbon dioxide for preparing $CO_2$-charged brine and injecting supercritical $CO_2$ ($ScCO_2$) into the porous sandstone samples. In path I, the core samples were kept at 10 MPa fluid pressure and 60 °C for 7 days. Path II continued by injecting dry $ScCO_2$ at 10 MPa injection pressure (constant flow regime, 20 ml/min) into the core sample until brine dry out.

### 2.4. Geomechanical triaxial compression

The intact and reacted BB and TR sandstones were tested in a servo-hydraulic operated system for triaxial rock mechanical measurements (AutoLab 2000 at AGH University Laboratory). The triaxial apparatus allows controlling pore pressure, confining pressure, and deviatoric stress on cylindrical samples independently with operational limits of 140, 140, and 200 MPa, respectively. In the triaxial apparatus, the maximum limit for axial loading is 1890 kN. The specimen is mounted in a sealed flexible membrane, preventing contact between the confining fluid and the specimen (here, oil was used). Two linearly variable displacement transducers (LVDTs) measured the sample deformation in the axial direction. Another LVDT sensor measured the radial deformation of the specimen. At each stress level, the vertical (or axial) strain was calculated by dividing the measured axial displacement over the initial height of the specimen. The horizontal (or radial) strain was computed by dividing the changes in the specimen's diameter over the specimen's initial diameter. For further details refer to Sharifi et al. (2021, 2023).

To maintain uniform saturation conditions, wherein complete dry-out and salt precipitation necessitate desiccation of the pore fluid, all three scenarios (intact, $CO_2$-acidified brine treated, and salt crystallization-affected) were examined under dry pore space conditions. Two sets of short-term loading cycles were carried out using the following test protocols:

A. non-destructive testing to determine Young's modulus and Poisson's ratio during the sample loading was repeated several times at three different temperatures 40, 60, and 80 °C for a given value of confining pressure (5, 10 and 15 MPa). Young's Modulus was determined as tangent modulus where the point of tangency falls within the linear range of the stress-strain curve, approximating inclination of linear elastic deformation. First, the confining pressure was set to a predefined value. The axial load was then controlled by changing the pressure exerted by the piston loading the sample. The amplitude of this load was 2 MPa. After reaching 2 MPa, the load was removed. A 0.1 MPa/s loading rate was adopted for the tests.

B. destructive tests where the confining pressure applied to the sample was set at 15 MPa. Then, the sample was loaded with the vertical stress exerted by the piston until the sample strength limit was reached. The test continued until the stress was 100% higher than the sample failure stress, and the load was then reduced. The axial displacement rate was 0.01 mm/s.

## 3. Results and Discussion

### 3.1. Sample characterization

Figure 1 presents optical images, x-ray micro-computed tomography, scanning electron microscopy, x-ray diffraction patterns of Boise Buff (BB) and Torrey Red (TR) sandstones' pore and matrix structures. It visualizes how different these two reservoir representative core samples are manifested in their respective sedimentary facies' classifications and reservoir quality.

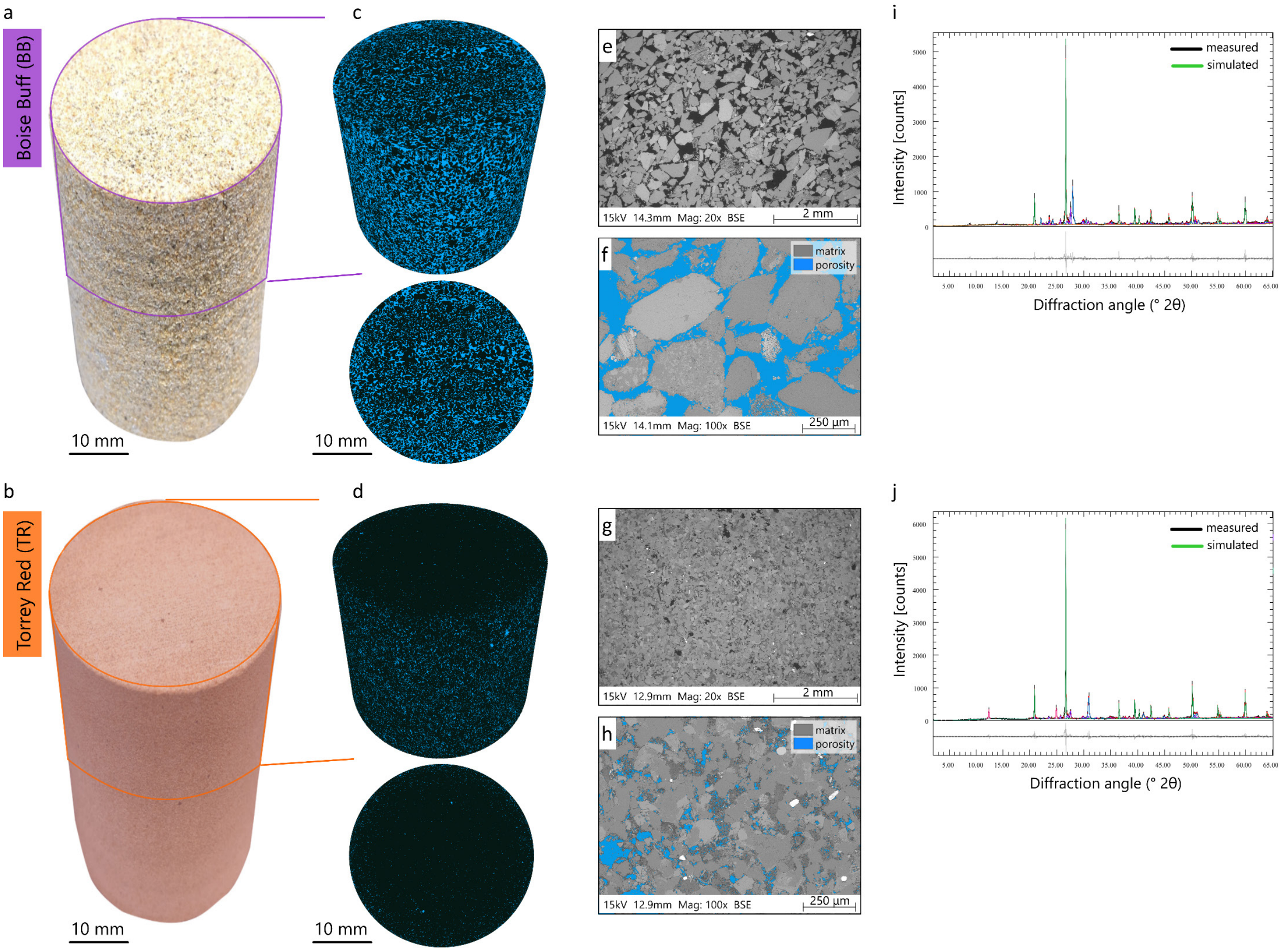

**Figure 1.** Sample characterization of tested (top) Boise Buff (BB) and (bottom) Torrey Red (TR) reservoir sandstones. (a-b) optical image, (c-d) segmented x-ray micro-computed tomography, (e-h) scanning electron microscopy, (i-j) x-ray diffraction pattern. The subplots visually convey the distinct properties of these two reservoir representative core samples, reflecting their sedimentary facies and petrophysical properties. The interpreted XRD patterns are given in Table 2.

There is a distinct difference between the pore and grain characteristics of the BB and TR sandstones. The BB sandstone shows coarser matrix grains, better sorting, less cementation, bigger intergranular porosity, and better pore connectivity than TR (Fig. 1). The amount of pore-filling material resulting in deteriorated percolating flow pathways is notably higher in TR sandstones. The smaller grain size of TRs results in smaller throat sizes, contributing to lower fluid permeability. The higher compaction level (combination of mechanical and chemical) suggests that TR sandstones are mechanically more competent than BBs.

The whole-rock (bulk) mineralogical composition of reservoir sandstones is presented in Table 2, showing that the BB and TR core samples consist mainly of quartz, feldspar (plagioclase and microcline) with minor amounts of clay, carbonate, and mica (muscovite). The major difference between the BBs and TRs is notably higher plagioclase (27%) with minor kaolinite clay content (2 %) in BBs. The TRs on the other hand contrasts with 12% kaolinite and approximately 8% dolomite content.

**Table 2.** Averaged mineralogical composition of Boise Buff (BB) and Torrey Red (TR) sandstones, based on XRD quantitative mineral analysis.

| | Quartz | Plagioclase | Microcline | Muscovite | Kaolinite | Dolomite | minor others |
|---|---|---|---|---|---|---|---|
| **Boise Buff (BB)** | 51 | 27 | 17 | 2 | 2 | 0 | < 1 |
| **Torrey Red (TR)** | 50 | 6 | 22 | 1 | 12 | 8 | < 1 |

### 3.2. Stress-strain relationship

Figure 2 presents the relationship between stress [MPa] and strain [milistrain] during triaxial compression experiments, in which the applied stress was gradually increased to measure the BBs and TRs deformation. Axial, radial, and volumetric strains are shown in the triaxial setting. The presented test results correspond to the continuous increase in applied stress until failure (fracture development). This provides the peak strength of the studied porous rocks. The stress-strain curves also determine the tentative elastic domain range and yield strength for the tested specimens.

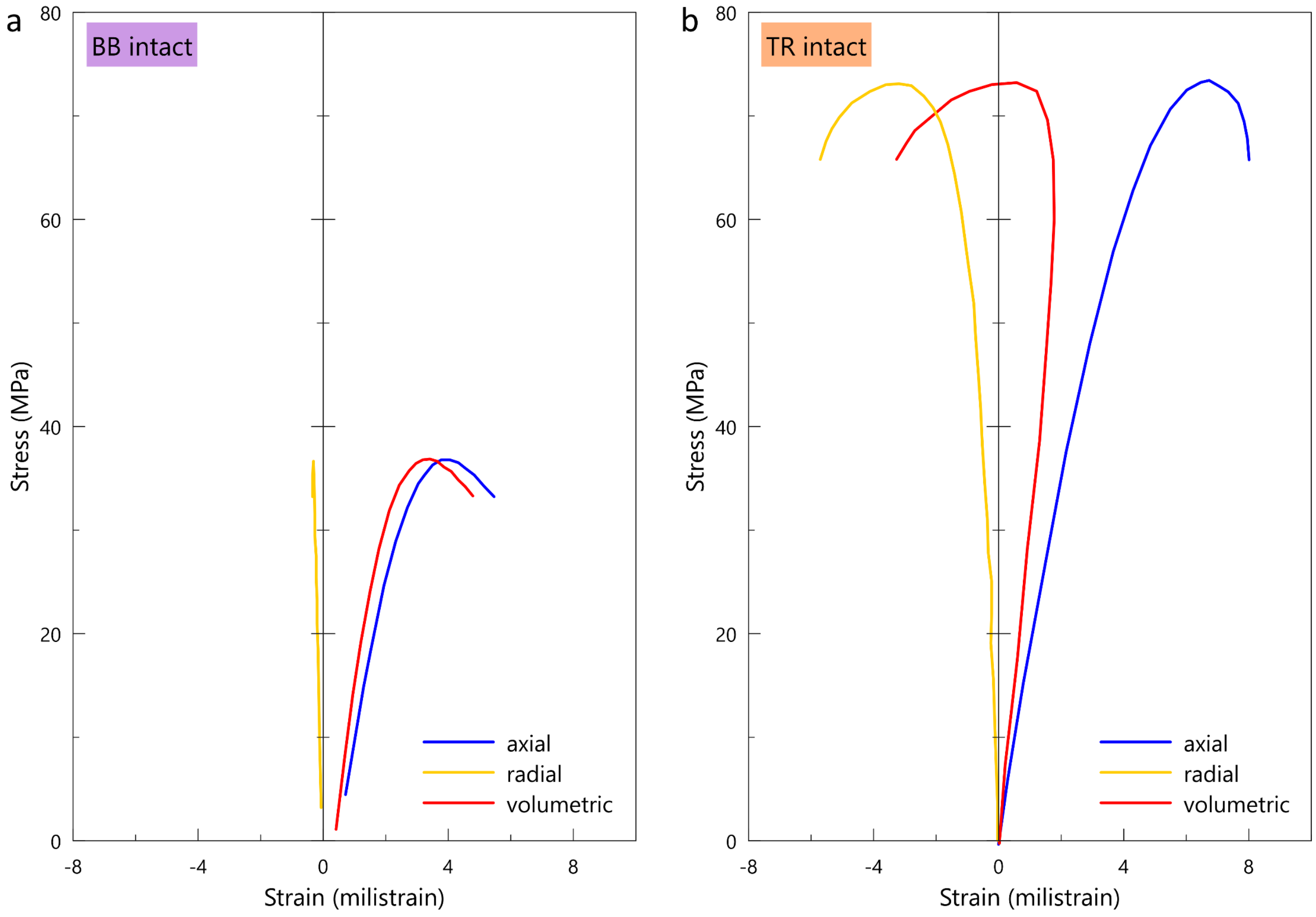

**Figure 2.** Stress [MPa] versus strain [milistrain] relationship during triaxial compression experiments on intact (a) Boise Buff (BB) and (b) Torrey Red (TR) sandstones. The subplots depict the development of axial, radial, and volumetric strains in the triaxial setting.

As subplots of Figure 2 demonstrate, the peak strength of the intact TR sandstones is approximately twice than that of intact BB core samples. The measured average compressive strength of BB and TR reservoir sandstones are approximately 31 and 77 MPa, respectively at 15 MPa confining pressure and 50°C. The less compacted (less cemented) BB reservoir rocks deform relatively faster, reach the yield strength at lower strain levels, and span a more constrained plastic domain before failure (rupture). They also show a very limited radial strain, as depicted in Figure 2a.

The cemented TR sandstones are characterized by notably high mechanical strength in the unreacted condition. The stress-strain curve in the elastic domain develops gently within the elastic deformation domain. The plastic domain is more extended compared to BBs, with a distinction in the plastic deformation signature (Fig. 2b).

## 3.3. Deformation behavior

### 3.3.1. Deformation of intact sandstones

Figure 3 presents crossplots of Young's modulus (E) [GPa] versus Poisson's ratio (ν) for intact (non-reacted) BB and TR sandstone core samples. The subplots are color-coded with experimental confining pressure [MPa] and test temperature [°C]. Figures SI-1/2 demonstrates laboratory E and ν values plotted for BB and TR individual specimens separately against changes in stress and temperature test conditions (refer to Supplementary Information).

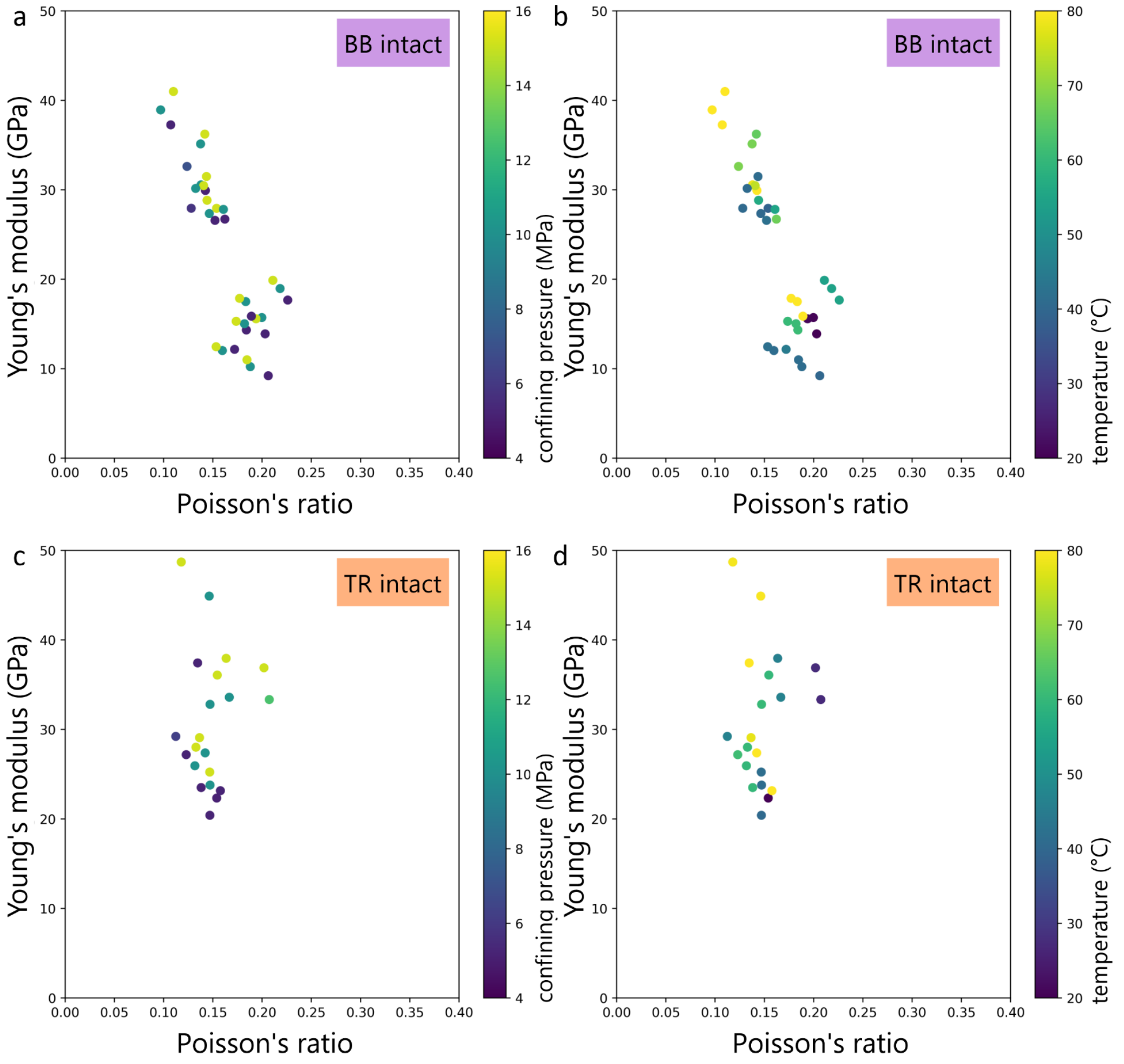

**Figure 3.** Crossplots of Young's modulus (E) [GPa] versus Poisson's ratio (ν) for intact (non-reacted) (top) Boise Buff (BB) and (bottom)Torrey Red (TR) porous sandstones. The subplots are color-coded with (a, c) experimental confining pressure [MPa] and (b, d) test temperature [°C].

Subfigures 3a-b reveal that the triaxial compression tests characterized two subgroups of intact BB sandstones, indicating natural heterogenies within the layered porous rocks. Group A shows E= 26-41 GPa and ν= 0.09-0.17, while the group B was characterized with E= 9-21 GPa and ν= 0.15-0.24. An increase in confining pressure (Pc) from 5 to 15 MPa during the experiments resulted in increase in Young's modulus for both subgroups (Fig. SI-1/2). An increase in the experimental temperature (T), ranging from 20-80 °C, resulted in increase in the elastic modulus (higher E values) (Figs. 3b and SI-2b). The increase in Pc caused a slight decrease in Poisson's ratio of both subgroups but the impact of temperature variations in Poisson's ratio is indistinct (Fig. SI-1/2).

Similarly, the elastic moduli of TR sandstones suggest two subgroups but with a smaller gap between the measured values (Figs. 3c-d). The E-ν ranges are as follows: In subgroup A, E= 18-29 GPa and ν= 0.11-0.16, and in group B, E=32-49 GPa and ν= 0.12-0.22. The impact of an increase in Pc and T on E-ν values of TR core samples are similar to BB, particularly for subgroup B (Fig. SI-1/2e-h).

As confining pressure rises, the rock undergoes a more even stress distribution, concurrently experiencing a reduction in deformability. This phenomenon is attributed to the increased confining pressure diminishing void spaces within the rock, rendering it less susceptible to deformations. Consequently, porous sandstones exhibit greater stiffness, exemplified by the observed increase in Young's modulus at elevated confining pressures of Figure 3. The elevation in confining pressure induces a reduction in pore spaces within the rock, emphasizing a direct link between increased confining pressure and enhanced mechanical integrity.

The influence of confining pressure extends to Poisson's ratio, primarily through its impact on the lateral contraction of the rock. The higher confining pressure effectively constrains lateral deformation, manifesting in a consequential decrease in Poisson's ratio. This observed decrease in Poisson's ratio bears significance, providing insights into how the mechanical behavior of the rock dynamically responds to varying stress conditions (Fig. 3). These interrelated effects collectively contribute to the intricate relationship between confining pressure, pore structure, and the mechanical properties of porous sandstones.

The influence of temperature involves a combination of effects, including thermal expansion of minerals and potential enhancement of grain-to-grain contact, among others. As temperatures elevate, individual mineral grains within the sandstone undergo thermal expansion, fostering a tighter packing that translates into a notable increase in stiffness or modulus. This thermal rise not only influences the physical dimensions of the mineral constituents but also fosters enhanced cohesion between grains, culminating in a rock matrix characterized by greater stiffness and reduced deformability.

It's crucial to acknowledge the nuanced nature of Young's modulus behavior with temperature in porous sandstones. The complexity arises from diverse factors, including mineral composition, porosity, cementation, and fluid type, each contributing to the material-specific interplay of these influences. The literature underscores multifaceted understanding of these interdependent factors, emphasizing their role in shaping the thermal response of porous sandstones (Orlander et al., 2021; Ranjith et al., 2012; Sun et al., 2016; Wang et al., 2022).

3.3.2. Deformation of acid-treated sandstones

The results of rock mechanical tests of BB and TR sandstones after treatment with $CO_2$-acidified brine (10 MPa fluid pressure and 60 °C for 7 days) are presented in Figures 4 and SI-3/4. The $CO_2$-reacted BB and TR sandstones show E moduli of 17-28 and 21-34 GPa, respectively. The ν ranges 0.13-0.23 and 0.18-0.33 for BB and TR. Therefore, the measured Poisson's ratio of these two sandstone sets are slightly higher than the non-reacted (intact) core samples, particularly evident for TR samples. For the tested BB and TR sandstones, the elastic modulus (E) systematically increases as the experimental temperature and confining pressure increase (Figs. 4 and SI-3/4). Increase in T resulted in an increase of ν in BB specimens, but a decrease in TR porous rocks (Figs., 4b,d and SI-4d,h). As Figure SI-4 shows, changes in the experimental stress regime (i.e., confining pressure) correspond with a minor increase in E and ν BBs but more pronounced growth in TRs.

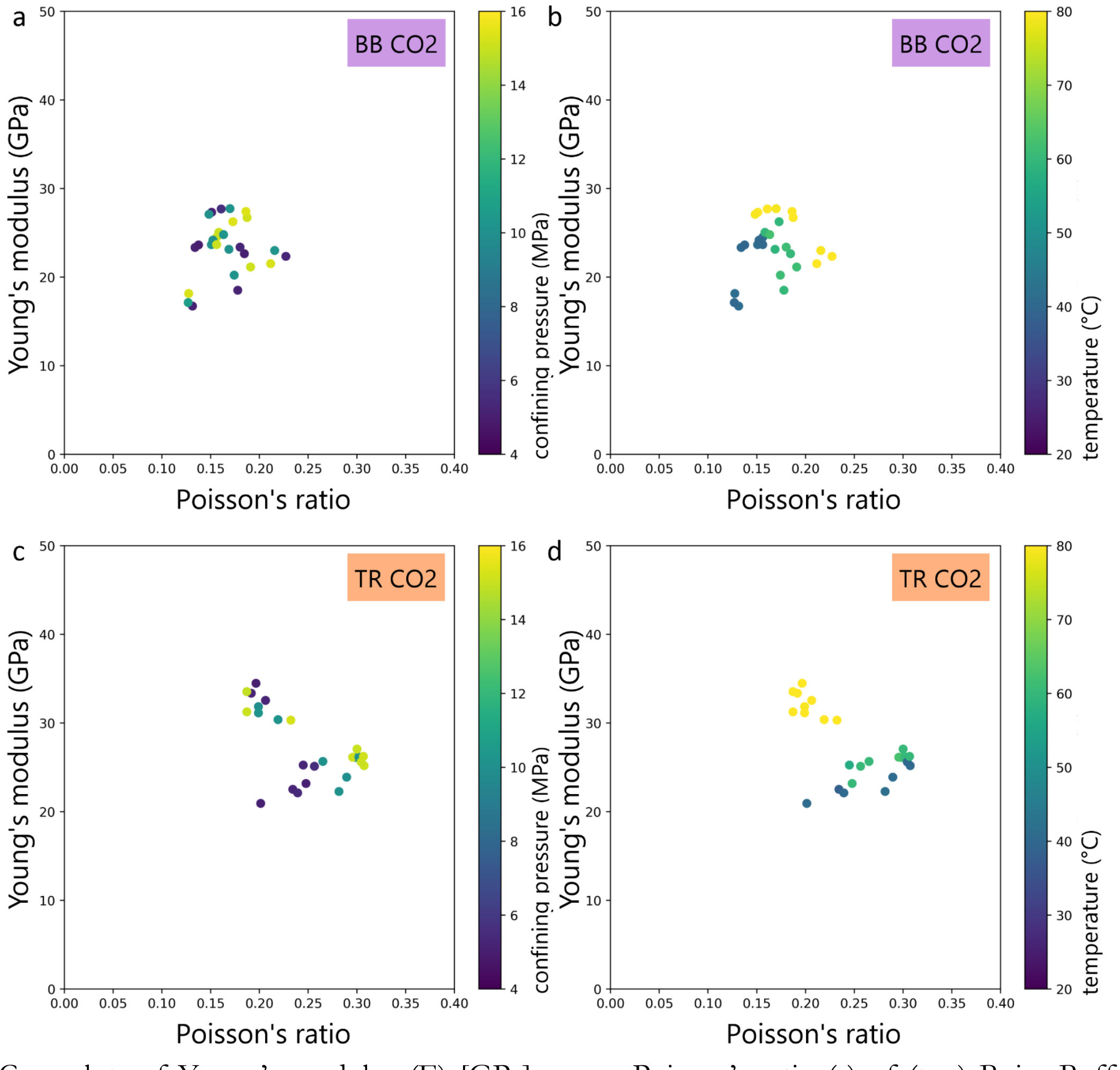

**Figure 4.** Crossplots of Young's modulus (E) [GPa] versus Poisson's ratio (ν) of (top) Boise Buff (BB) and (bottom)Torrey Red (TR) sandstones after treatment with $CO_2$-acidified brine. The subplots are color-coded with (a, c) experimental confining pressure [MPa] and (b, d) test temperature [°C].

The acidification of the brine, facilitated by continuous supply of $ScCO_2$, may have initiated subtle adjustments within the sandstone grain contacts and porous matrix, influencing rock microstructure and grain contacts without entailing any rapid or major mineralogical transformations. Concurrently, the treatment might have subtly influenced the pore structure of the rock, impacting the elastic modulus and Poisson's ratio of the sandstone. The observed elevation in elastic modulus with experimental temperature and confining pressure implies the influence of these external factors on the sandstone's mechanical response. The varied responses in TR and BB sandstones likely stem from differences in their initial mineralogical composition, cementation and fines, and pore volume compliance, resulting in distinctive changes. These findings provide insights into the effects of acidified brine and clean reservoir sandstone, shedding light on the complex mechanical behavior of $CO_2$ storage reservoir rocks. For a more in-depth exploration, we direct the reader to Table 1 (see appendix) and the detailed literature review presented in the introduction, providing an overview of chemo-mechanical interactions and range of variations in geomechanical properties in these conditions.

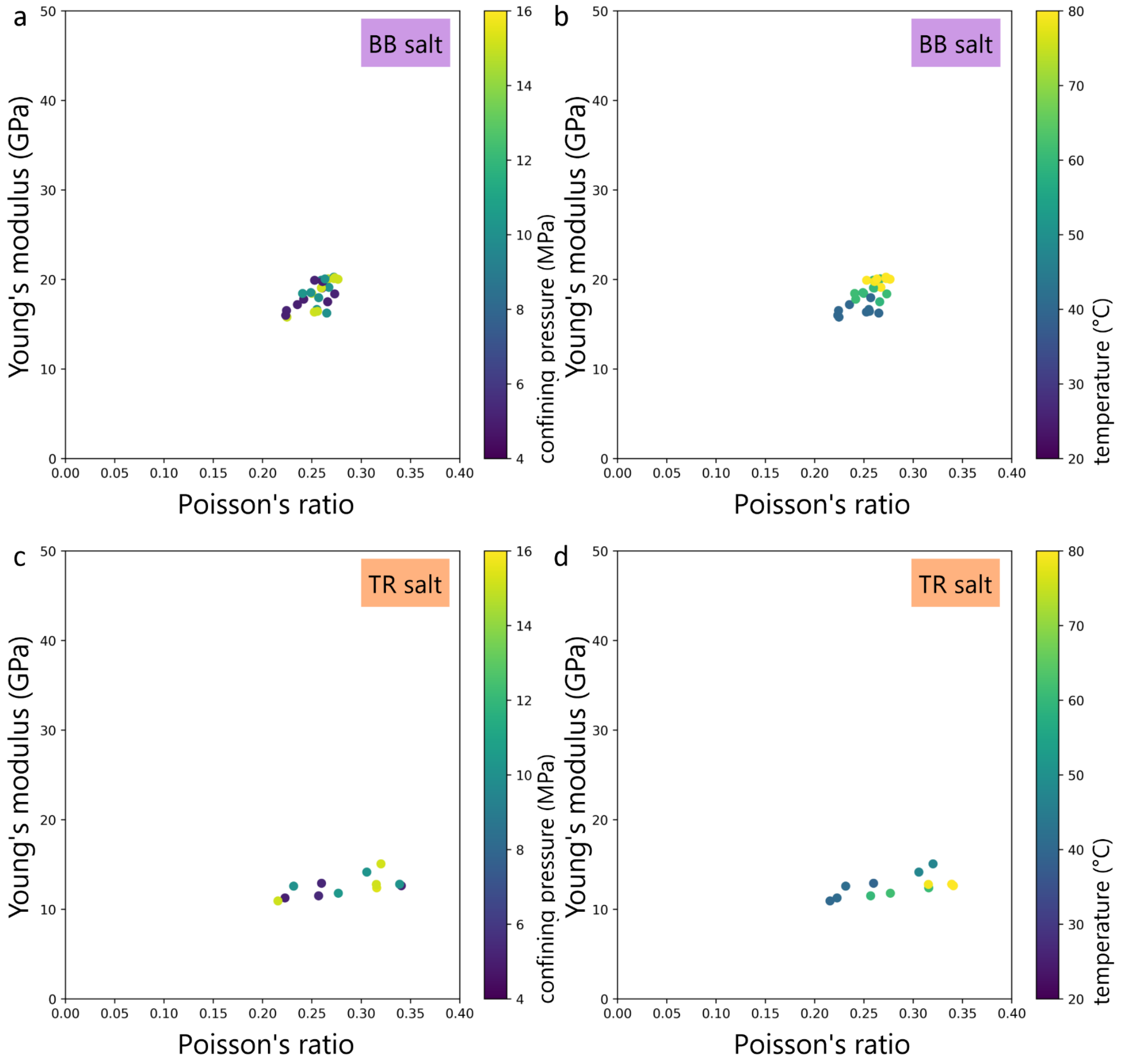

**Figure 5.** Crossplots of Young's modulus (E) [GPa] versus Poisson's ratio (ν) of (top) Boise Buff (BB) and (bottom)Torrey Red (TR) sandstones affected by $CO_2$-induced salt precipitation. The subplots are color-coded with (a, c) experimental confining pressure [MPa] and (b, d) test temperature [°C].

### 3.3.3. Deformation of crystallization-affected sandstones

Figures 5 and SI-5/6 show the elastic parameters of sandstones affected by $CO_2$-induced salt crystallization. A marked reduction in Young's modulus is evident in both BB and TR sandstones, with Young's modulus ranging from 15-21 GPa for BB and 10-15 GPa for TR core samples. These values represent approximately half of the average values measured for intact and CO2-reacted specimens, underscoring the substantial impact of salt crystallization on the mechanical properties. The TR core samples, characterized by tighter properties such as lower porosity and significantly reduced permeability, exhibit a more pronounced decline in elastic moduli compared to BBs. This observation suggests that the initial characteristics of the sandstone play a crucial role in determining the extent of mechanical degradation induced by salt crystallization. The measured Poisson's ratios show a noteworthy increase of approximately 1.5 compared to previous cases, covering a range of 0.22-0.29 for BB and 0.21-0.34 for TR sandstones. Subplots in Figures SI-5/6 reveal that the increase in confining pressure and test temperature during the experiments leads to an elevation in both E and ν parameters, the slope of which might seem gentler than previous cases.

It is worth mentioning, one of the TR core samples were damaged rapidly and intensively during the initiation phase of the experiments with salt-damaged TRs, indicating a dramatic decline in the strength parameters after the salt crystallization in porous media and continued growth. That is the reason why Figure SI-6 includes only two plotted samples.

### 3.4. Mechanical parameters deterioration

Figure 6 juxtaposes Young's and shear moduli of intact, treated with $CO_2$-acidified brine, and salt crystallization-damaged BB and TR sandstones as a function of increased confining pressure. It shows how the mechanical properties of these three sets of experiments were affected as stress regimes and geochemical factors varied.

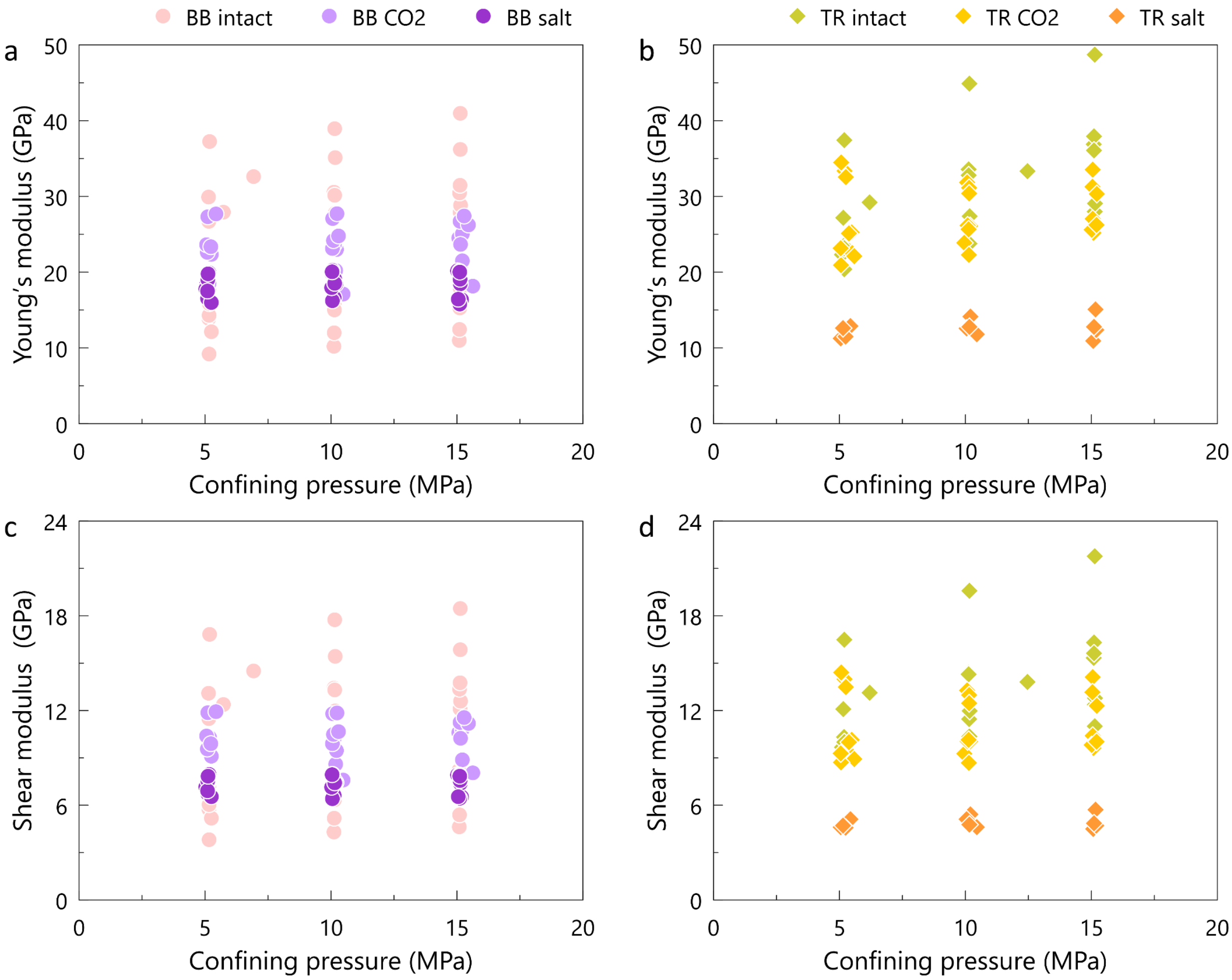

**Figure 6.** Stress sensitivity of (a, c) Young's [GPa] and (b, d) shear moduli [GPa] of intact, $CO_2$-acidified brine-treated, and salt crystallization-damaged (left) Boise Buff (BB) and (right) Torrey Red (TR) sandstones as a function of increase in confining pressure [MPa]. The subplots illustrate the impact on the mechanical properties of these three sets of experiments, showcasing variations in stress regimes and geochemical factors.

As Figure 6 illustrates for tested subgroups of BB and TR rocks, the mechanical properties of a given sandstone rock type are intricately shaped by a myriad of natural heterogeneities, encompassing variations in pore volume, grain framework, mineral composition, grain contacts, and cementation, among others. The inherent heterogeneity, rooted in complex sedimentation processes, manifests as discernible differences in grain size, sorting, and layering within the rock. Simultaneously, the mineralogical composition, incorporating minerals such as quartz, feldspar, and clay, introduces an additional layer of complexity to the sandstone's mechanical behavior. Rock properties are then subsequently controlled by the degree of mechanical and chemical compactions and burial history (Bjørlykke, 2015; Bjørlykke and Jahren, 2015; Nooraiepour, 2022, 2019, 2018), reflected in present-day porosity-permeability profile of these two reservoir quality classes (refer to Section 3.1).

These compositional variations play a pivotal role in determining the elastic moduli (and predictive modeling of long-term subsurface coupled THMC processes), encompassing parameters like Young's modulus and shear modulus, as well as influencing strength parameters such as compressive and tensile strength. Additionally, these variations govern the rock's deformation behavior, contributing to localized differences in brittleness and ductility. Anisotropy, stemming from grain orientation, imparts directional dependence to the mechanical properties, further enriching the understanding of the sandstones' response to applied stress.

By juxtaposing datapoints in Figure 6, two discernible trends emerge, each contributing to a deeper understanding of the mechanical behavior of sandstone classes (BB or TR). Firstly, as defined by inherent heterogeneities, the initial variation domains depict distinct differences between the two classes, with BB exhibiting a broader scatter than TR, as also illustrated in Figure 3. It is essential to emphasize that each class is characterized by a range of properties rather than a single estimator. However, as we transition from intact to $CO_2$-treated and salt-damaged rocks, the degree of variation across the ordinate (y-axis representing moduli) diminishes.

Secondly, the deterioration of Young's and shear moduli due to $CO_2$-induced salt crystallization damage is notable. While each experimental subsets indicate the expected trend as confining experiments is increased in triaxial experiments (in addition to temperature change), a systematic decline is observable. While this impact is not distinct enough due to the wide spread of intact BB specimens, the marked strength reduction of TRs to approximately half is demonstrated in Figs. 6b,d. The post-interaction TR specimens, affected by salt crystallization damage, vividly portray a potential upper-band mechanical weakening, clearly depicting the deleterious effects of salt-induced alterations on the sandstone's mechanical integrity. We posit that the observed decline represents the upper limit of damage, manifested through the continuation to complete dry-out and extensive precipitation at our experimental conditions. Notably, our experimental observations align with findings documenting such a decline in dynamic elastic parameters, as computed using measured compressional and shear wave velocities (Zhang et al., 2020).

The tighter TR rocks (with lower reservoir quality) demonstrate higher Young's and shear moduli in addition to the higher peak strength compared to more porous-permeable BB specimens (with higher reservoir quality). However, salt crystallization and crystal growth inside the porous sandstone geometries damaged the TR sandstones more intensively than the BBs, leading to more aggressive mechanical weakening (Fig. 6d). Computed shear moduli show a decrease from a maximum of 22 and 18 GPa for intact sandstones to 6-8 and 4-6 GPa for salt-damaged TR and BB rocks, respectively (Fig. 6). As an average, the decrease for BBs is to half and for TRs to a third. The $CO_2$-reacted specimens show $\mu$ values between these two extremes. The significant reduction in rigidity observed here indicates an increased susceptibility to shear failure. This may occur as a potential risk when the in-situ shear stress surpasses the shear strength of CO2 storage reservoirs, particularly in the near-wellbore regions.

To evaluate the brittleness and ductility of the intact and reacted sandstones, we constructed crossplots of Young's modulus versus Poisson's ratio and shear modulus or rigidity (μ) versus incompressibility (λ) in Figure 7. Figure 7b is a modification of the well-known λϱ-μϱ rock physics crossplot, where the bulk density (ϱ) is factored out. The classification of regions is adopted from review of literature on brightness-ductility classifications presented in Nooraiepour et al. (2017). Figure SI-7 shows Lamé's first and second parameters (λ-μ) plots for different experimental cases separately.

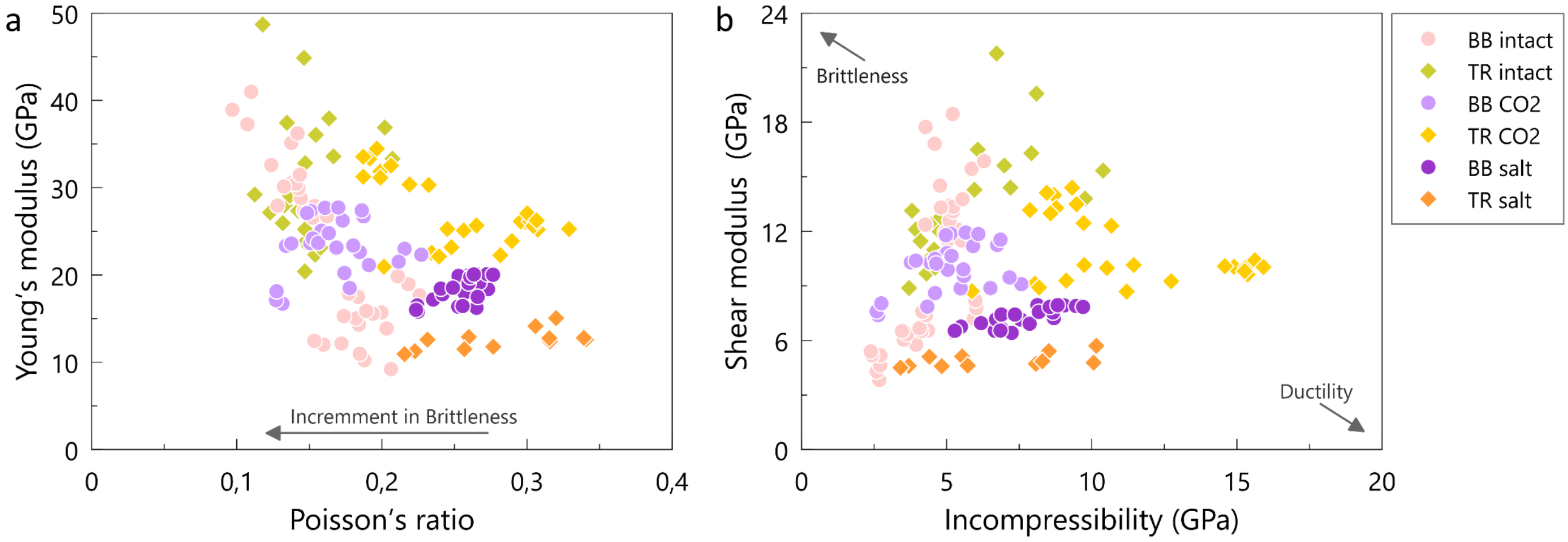

**Figure 7.** Rock physics crossplots evaluating brittleness and ductility of intact and reacted sandstones: (a) Young's modulus vs. Poisson's ratio, and (b) shear modulus vs. incompressibility. Subplot b is modified from λϱ-μϱ crossplot, where bulk density (ϱ) is factored out. Regional classification follows the literature review on brightness-ductility presented in Nooraiepour et al. (2017).

Comparison between Figs. 7a-b demonstrates that different proxies for brittleness evaluation can provide slightly different assessment results. Overall, Figure 7 suggests that the treatment of samples caused a shift in the geomechanical properties of BB and TR sandstones toward the more ductile regions. In both subplots, to a large extent, a movement of data points toward the bottom right is evident, particularly for the E-ν crossplot (Fig. 7a), when one juxtaposes static rock mechanical parameters of intact, $CO_2$-reacted and salt-affected porous rocks. However, the λ-μ relationship of tested specimens does not necessarily suggest that reacted BB and TR show markedly increased ductility (Fig. 7b).

## 4. Salt crystallization damage in the carbon sequestration context

### 4.1. Mechanical weakening of saline aquifers

As presented in Table 1 (see the appendix), laboratory investigations to comprehend the interactions between $CO_2$, brine (water), and rock constituents have provided extensive insights into relatively short-term geochemical consequences, typically spanning a maximum of 2-3 years. However, there still exists a knowledge gap concerning the protracted reaction mechanisms and combined consequences when the contribution of quick carbonate (calcite) mineral dissolution is minuscule, particularly in clean reservoir sandstones. Experiences from industrial and pilot CCS projects worldwide have undoubtedly built solid scientific and operational grounds. However, laboratory and numerical efforts have repeatedly shown the necessity of further thorough investigation of physico-chemical processes with potential coupled THMC challenges for scale-up.

It has been shown that $CO_2$-brine-rock interactions are ubiquitous in carbon storage reservoirs with the potential for triggering mechanical damage in sandstone rocks. These mechanical damages primarily result from physical changes in the rock microstructure due to fluid-rock reactions/interactions through mechanisms such as mineral alteration/transformation, clay swelling and shrinking, pressure buildup in pores, redistribution of fines, thermal stress, and local stress concentration within porous geometries (Table 1).

We expect the highest probability of (geo)chemically induced mechanical damages to occur in the near wellbore environment where the highest acidity (lowest pH) of formation water is achieved due to carbon dioxide dissolution and carbonic acid generation. This region also endures the fastest $CO_2$ flow through the pore volume, highest viscous forces, potentially highest pressure buildup, lowest chemical buffering of reactive pore fluid, and most significant thermal stresses. Additionally, $CO_2$-induced salt precipitation may substantially threaten the mechanical integrity of near injection well zones.

The study of salt crystallization has seen a dynamic convergence of various scientific disciplines to unravel the microscopic underpinnings of processes governing salt precipitation dynamics, salt weathering mechanisms, and consequent damages. An expression to calculate the crystallization pressure as a function of the supersaturation was first derived by Correns and Steinborn (Correns and Steinborn, 1939, translated and commented upon by Flatt et al., 2007). Everett (Everett, 1961) pursued an alternative method to address the issue, establishing a connection between crystallization pressure and the characteristics of curved interfaces existing between crystals and their respective solutions. In the last two decades, consensus has been reached that salt precipitation and growth under confined and supersaturated conditions exert crystallization pressure.

The prevailing theory posits that crystallization pressure within pores arises from a thermodynamic interaction resulting from a non-equilibrium state created by solution supersaturation (Flatt, 2002; Flatt et al., 2007; Scherer, 2004; Steiger, 2005a), and interfacial energies as repulsive disjoining force operating between the pore wall and the burgeoning crystal (when there is a liquid film between the crystal and the confining surface) (Coussy, 2006; Espinosa-Marzal and Scherer, 2010; Hamilton et al., 2010; Scherer, 2004; Steiger, 2005b). A formulation for crystallization pressure, harmonizing with both models, was subsequently devised, which incorporates Pitzer's (Pitzer, 1991) ion interaction approach to account for the nonideal behavior of the liquid phase, ensuring comprehensive consideration (Flatt et al., 2017; Flatt, 2002; Steiger, 2005a, 2005b). The theoretical formulations signify the maximum pressure applied to a pore wall, achievable only at equilibrium when a sizable crystal is confined within a pore with narrow throats. It also marks the highest transient pressure exerted when a crystal growing in a large pore initially encounters the wall. As the crystal expands, it diminishes the local supersaturation, leading to a subsequent decrease in pressure (Flatt et al., 2017) under the assumption that crystal growth consumes the solutes faster than its transport to the sites by mechanisms such as diffusion, capillary backflow, and water film flow across grain surfaces.

The growing crystal experiences distinct pressures on its various faces. Faces in contact with the liquid solution undergo the pressure of the pore solution, whereas those in contact with the confining surface via the liquid film are presumed to experience crystal pressure. The existence of this disjoining force is attributed to the fact that the energy associated with the salt or mineral interface, which would form when the crystal grows adjacent to the pore wall, exceeds the cumulative energies of the salt-to-solution and mineral-to-solution interfaces (Espinosa-Marzal and Scherer, 2010; Flatt et al., 2017). The crystal pressure is determined to be higher than that of the surrounding solution, as indicated by Scherer et al. (2001) and Steiger (2005a,b). Therefore, the salt crystallization pressure will apply as crystals form within their porous hosts, placing the hosts under tensile stresses (Derluyn et al., 2014; Flatt et al., 2014). These forces can harm materials like reservoir rocks with relatively low tensile strength.

Steiger (2005a) computed (maximum) crystallization pressure (using Eq. 19 in his paper) in supersaturated solutions containing aqueous NaCl, $NaNO_3$, $Na_2SO_4$, and $MgSO_4$, commonly found in natural rock environments. The calculations incorporated both the anhydrous and hydrous phases of these salts. The results demonstrated that the crystallization pressures generated at low to moderate supersaturation levels are capable of inducing tensile stresses on the pore walls. Notably, these stresses surpass the tensile strength of many saline aquifer reservoir rocks. For instance, at supersaturation levels of 1.2, all the salts, except mirabilite, exhibited significantly higher crystallization pressures than the tensile strength of sandstone (assuming approximate average of 3-4 MPa) (Shao et al., 2022; Steiger, 2005a).

The significance of thin liquid films in crystallization pressure has already been highlighted in theory (Desarnaud et al., 2016; Espinosa-Marzal and Scherer, 2010). To generate crystallization pressure, upholding a liquid film through repulsive forces acting between the crystal and the wall is imperative to facilitate the ongoing growth of the crystal by incorporating additional ions into its lattice. It is crucial to note that if a growing crystal completely spans the void between two enclosing walls, no crystallization pressure can emerge, as there is no opportunity to add extra salt layers and create such pressure. It has been shown (experimentally and numerically) that salt crystals can draw water from neighboring mineral grains by establishing persistent water films covering/wetting mineral grains (Masoudi et al., 2023; Miri et al., 2015; Nooraiepour et al., 2018; Nooraiepour et al., 2019; Ott et al., 2021; Qazi et al., 2019), primarily due to surface energy influences. Consequently, a perpetually replenished brine source is made available to sustain the advancing evaporation front, ensuring a consistent expansion of salt crystals upon pre-existing salt aggregates. The water films may exhibit notable mobility and conductivity for solute transport. The significance of the presence of water film was studied using hydrophilic and hydrophobic glass plates (Desarnaud et al., 2016). Hydrophobic glasses, which impede the formation of a wetting film, did not register any pressure. In contrast, hydrophilic glass slides facilitated the formation of a thin film, resulting in a significant crystallization pressure. The crystallization pressure for halite (NaCl) and sylvite (KCl) increased with higher salt concentrations, aligning with theoretical expectations. For a given salt concentration, halite exhibited markedly higher crystallization pressure (Desarnaud et al., 2016).

### 4.2. Conceptual model of microscale crystallization damage

Figure 8 provides scanning electron microscopy (SEM) of microscale adverse effects of salt crystallization in reservoir sandstones. We use these direct observations, describe them below, and in Figure 9 propose a series of conceptual damage mechanisms induced by $CO_2$-induced salt precipitation. We limit the presentation of direct imaging evidence to SEM micrographs in this paper, avoiding micro-CT segmented images to prevent unwarranted speculation and biased interpretations.

The imaging specifications hinder the detection of cracks and fissures in individual matrix grains of brittle quartz and feldspar mineral grains. Our exploration of the Digital Rock Portal reveals that only high-resolution (super-resolution) tomography can partially capture certain features illustrated in Figure 8 and conceptualized in Figure 9.

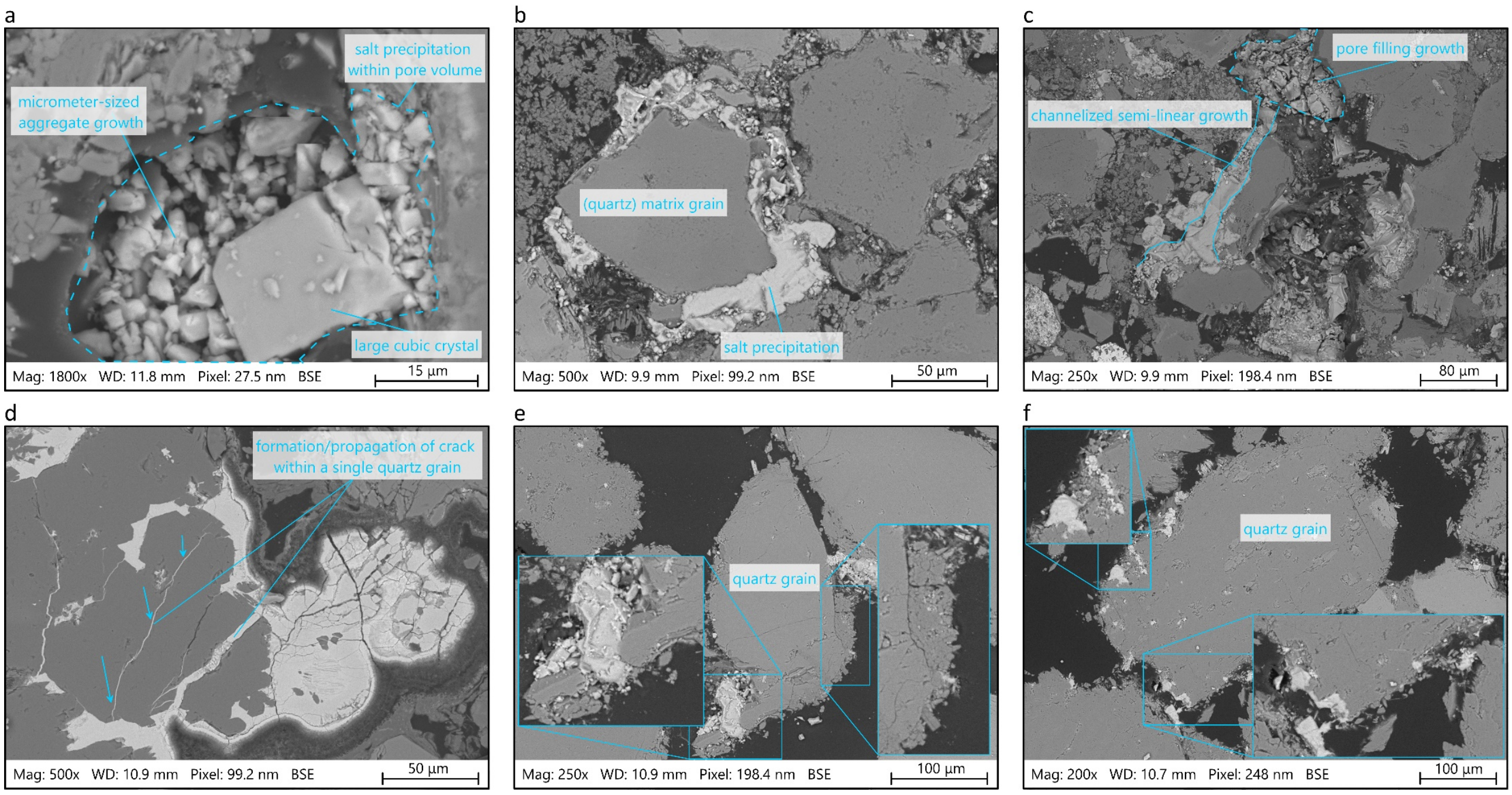

**Figure 8.** Scanning electron microscopy (SEM) images showcasing salt crystallization effects on sandstones: (a) salt precipitation and continuous growth in a pore body, featuring a large cubic crystal and micrometer-sized aggregated growth; (b) salt precipitation occupying throats between matrix grains and partially filling pore volumes around a quartz grain; (c) channelized semi-linear salt growth along several grains connected to a nearly complete clogged of pore volumes; (d) crystallization of salt bodies within cracks inside the quartz grain, with potential for evolution into significant fractures of varying sizes; (e-f) a quartz grain displaying topological damage features such as chipping and abrasion due to peripheral salt nucleation and growth.

Despite the limitations in spatial resolution of our tomography, conducted x-ray imaging on postreaction specimens successfully identified certain larger-scale conceptualized mechanisms. Notably, we observed dilations in various regions of the porous media. Considering the limited geochemical reactivity of our clean BB and TR sandstones, we interpret the expansion of samples as a consequence of localized volumetric increases induced by salt precipitation and growth (similarly reported also by Falcon-Suarez et al., 2020). If our ongoing investigation confirms the observation of rock frame dilation due to $CO_2$-induced salt precipitation in sandstones with high porosity-permeability and low reactivity to $CO_2$, then early detection of salt precipitation near the wellbore region becomes imperative. This is crucial for preserving both the injectivity properties of the reservoir and its mechanical integrity.

Upon initiating $CO_2$ injection, a distinct two-phase-flow zone emerges, featuring an aqueous phase alongside a $CO_2$-rich phase (Fig. 9a). This process, known as two-phase displacement, is primarily characterized by viscous displacement sweeping brine from around the injection well. As the flooding front, representing a shock front, progresses it leaves behind a zone where residual brine is confined in diverse configurations, such as thin wetting films enveloping grain surfaces and liquid bridges or pools within pores. This drained region faces a continuous flow of dry $scCO_2$ with low water vapor pressure, initiating an evaporation regime. Persistent flow leads to substantial water evaporation into the $CO_2$ stream, inducing dry-out. While both two-phase displacement and evaporation contribute to water removal near the well, the time scales of these processes remain distinct.

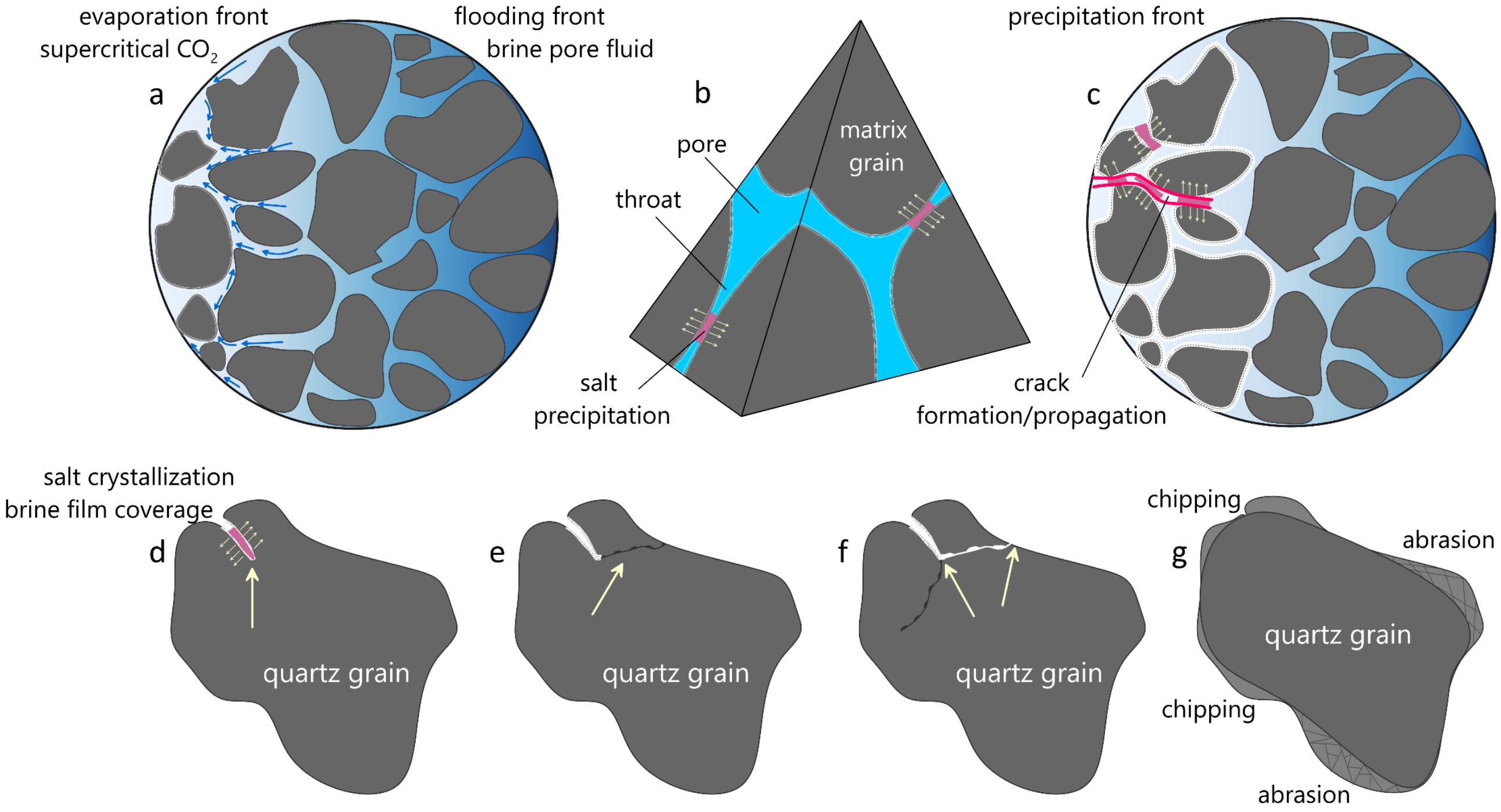

**Figure 9.** A schematic representation elucidating microscale mechanisms involved in salt crystallization damage in porous rocks. (a) depiction of evaporation front formation, flooding front, initiation of salt nucleation and growth, and capillary-driven backflow of water films. (b) salt crystal growth along grain surfaces, within the throats, and inside the pore bodies, exerting significant internal stresses on the surrounding matrix grains or porous medium's structure. (c) initial fissures, whether induced in the previous step or naturally occurring in the rock framework, are filled with brine, subject to nucleation and growth, and exposed to crystallization damage. This sequence of events extends to brittle sandstone grains with natural defects and initial cracks formed during transport and diagenesis (mechanical compaction). (d) fracture extension occurs at existing cracks within or around the edges of brittle grain matrix due to migrating water films and crystallization-induced mechanical stress. (e-f) additional cracks are initiated through further nucleation and precipitation events on a secondary substrate (previously crystallized salts), ultimately resulting in extensive fracturing and breakdown of the quartz grain. (g) nucleation and growth of salt crystals may also trigger topological grain damages such as chipping and abrasion.

The primary water mass exchange occurs in the dry-out zone, generating a saturation gradient across the drying front significantly greater than that in a pure viscous displacement. This gradient, driven by capillary pressure, drives water towards the evaporation front, intensifying evaporation (blue arrows in Fig. 9a). Additionally, as water evaporates into the $CO_2$ stream, increasing salt concentration in trapped brine leads to outward salt diffusion from the drying front. The distance between the dry-out and flooding fronts is primarily regulated by capillary backflow and solute diffusion. Once salt concentration reaches its solubility limit due to evaporation, salt precipitates, exhibiting a strong affinity for brine and facilitating further precipitation (Figs. 8-9/b-c). Salt-induced capillary backflow enhances the stability of water films, augmenting continued accelerated salt precipitation (Miri and Hellevang, 2016).

Continuous (repetitive) wetting-drying cycles, driven by capillary suction and continuous brine films supply in pore volumes, lead to crystal growth along grain surfaces, throats, and pore bodies (Figs. 8-9/b-c). As the salts crystallize inside the porous medium, they undergo a change in physical state from a dissolved solute to a solid crystal lattice, and crystallization pressures are built up. This induces considerable internal stresses on the surrounding matrix grains or porous medium's structure (Figs. 8-9/b-c). The internal stresses generated by salt crystallization can reach a point where they exceed the material's tensile strength. The tendency for crystals to preferentially grow in narrow pore throats (Miri et al., 2015) significantly influences the sandstone's internal geometric structure evolution, with narrow throats being particularly susceptible to salt damage (Figs. 8-9b).

Initial fissures, whether induced or naturally occurring, are filled with brine and exposed to crystallization damage. As crystals grow during drying, they exert substantial crystallization pressure on fissure walls, causing cracks (Figs. 8-9c). Small microcracks may initiate within the material, typically at weak points or grain boundaries that mark the crack initiation stage. The continued crystallization and growth lead to stress accumulation, during which these microcracks can propagate further into the material, extending and deepening the damage (Figs. 8-9c). During crack propagation, small pores are preferentially cracked and connected. If the in-situ stress continues to build, the precipitation-induced/-enhanced cracks may have the potential to evolve into more significant fractures (with various sizes). These micro- to macro-scale deterioration chain of events in material properties may sum up to colossal weakening of the solid and porous structure.

It should be noted that different salt solutions exhibit distinct crystallization characteristics, leading to varying degrees of rock deterioration (ongoing study). Thus, further research is recommended to explore the impact of different salt solutions on rock deterioration, focusing on crystal growth features.

Quartz and feldspar, as prevalent constituents of sandstone, exhibit pronounced brittle characteristics owing to their mineral structure and mechanical properties. The crystalline structure of quartz, comprising tightly packed silicon dioxide units, renders it susceptible to fracture and breakage under applied stress, particularly in the presence of mechanical discontinuities or stress concentrations. Similarly, feldspar, a framework aluminosilicate mineral, exhibits brittleness due to its complex lattice structure and susceptibility to cleavage along specific crystallographic planes. The brittle nature of quartz and feldspar influences the overall mechanical behavior of sandstone, contributing to fracture initiation, propagation, and the overall deformation response of the rock under varying stress conditions.

Figures 9d-f present a schematic illustration of a potential mechanical deterioration mechanism focusing on the disintegration process of quartz grains. This representation is derived from a meticulous examination using high-resolution SEM of BB and TR samples at the matrix grain scale, as depicted in Figures 8d-f. Building upon the earlier-discussed chain of events responsible for pore-scale mechanical damages, we extend the implications to brittle sandstone grains characterized by natural defects and initial cracks formed during transport and diagenesis.

Theoretically, water-filled pore bodies induce internal strain through volume expansion and mechanical stress during salt growth. These stress factors are initiated at point defects and vary in intensity based on the presence of inclusion trails. Fracture extension occurs at existing fractures from inclusion trails and subgrain boundaries due to migrating water and crystallization-induced mechanical stress (Fig. 9d). Additional fractures are initiated through further nucleation and precipitation events on a secondary substrate (previously crystallized salts) (Masoudi et al., 2022; Nooraiepour et al., 2021a) (Figs. 8c, 9e). Ultimately, extensive fracturing leads to the breakdown of the quartz grain (Figs. 8c, 9f).

Repetitive observations in SEM micrographs suggest that particles with topographical features or angular surfaces may undergo chipping breakage, where the application of load on an irregular particle result in the breakage of rough corners (Fig. 8e). This chipping phenomenon occurs when edges or corners of a particle break off. Additionally, attrition loading can induce abrasion, which occurs when the applied energy is insufficient to cause significant particle breakage. Abrasion is also characterized by the generation of fines around matrix grains with no noticeable change in the mean diameter of the feed particle (Fig. 8f).

Juxtaposing the results of two reservoir sandstone classes in this study clearly indicates that the significance of pore size (distribution) in the damage resulting from salt subflorescence (i.e., precipitation and growth inside the porous material) cannot be overstated. It profoundly impacts the crystallization pressure and the distribution of minerals (Derluyn et al., 2014; Nadelman and Kurtis, 2019; Shokri-Kuehni et al., 2018; Yu and Oguchi, 2010). In reservoir rocks characterized by smaller pores, the crystallization pressure is notably higher than in layers with larger pores, and precipitation occurs deeper inside the evaporation front. These two factors collectively contribute to increased deterioration and reduced mechanical strength to subsequent salt crystallization. And finally, salt crystallization is a cyclic process (Balboni et al., 2011; Desarnaud et al., 2013). Each wetting and drying cycle leads to additional salt crystallization, causing further expansion, stress, and damage. The cumulative effect can be substantial, particularly in structures and geological formations exposed to these cycles for extended periods.

## 5. Conclusions

New laboratory experiments provide crucial insights into the geomechanical response of two distinct sandstone reservoirs, Boise Buff (BB) and Torrey Red (TR), under varying geological and geochemical conditions. BB and TR sandstones exhibit notable variations in their sedimentary characteristics. BB sandstone represents better reservoir quality with coarser matrix grains, lower cementation, larger intergranular porosity, and improved pore connectivity, rendering it highly permeable. Conversely, TR sandstones feature finer matrix grains, tighter pore space, and narrower throat sizes but have greater mechanical strength. The main findings can be succinctly outlined as follows:

- Analysis of existing research reveals that the rapid reaction rate of calcium carbonates can explain the interactions of $CO_2$ and carbonic acid with reservoir sandstones containing calcite. However, certain sandstone reservoirs lacking fast-reacting minerals exhibit a decline in mechanical properties when exposed to $CO_2$ and $CO_2$-acidified brine. While some studies attribute this to alterations in grain-to-grain contact points, others merely report the observed changes. Concluding our research, we highlight an existing knowledge gap in this domain, underscoring the need for additional targeted studies to explicitly explain potential microscale alterations. It is imperative to delve into the mechanisms and specific governing factors, supported by microscale evidence, that elucidate conditions when geochemical reactions resulting in composition changes are absent, yet significant mechanical consequences are observable in clean reservoir aquifers.
- Stress-Strain Relationships: Triaxial compression tests reveal that TR sandstones exhibit nearly double the peak strength of BB samples. BB porous rocks reach yield strength at lower strains and exhibit a more confined plastic deformation region. In contrast, TR sandstones display higher mechanical strength, a stiffer characteristic, and a more extensive plastic deformation range.

- Geomechanical Alterations: $CO_2$-brine-rock interactions emerge as a critical factor affecting the rock mechanical properties of sandstones. The study demonstrates how changes in confining pressure and temperature affect the elastic properties of both sandstone types. Both sets exhibit distinct subgroups with varying Young's modulus (E) and Poisson's ratio (ν) values, highlighting the impact of natural heterogeneities on overall rock strength. Exposure to $CO_2$-acidified brine results in slightly reduced mechanical parameters compared to their intact counterparts.
- Salt crystallization stands out as a significant contributor to sandstone mechanical damage and weakening. The salt-affected specimens show a 50% reduction in Young's and shear moduli and twice an increase in Poisson's ratio compared to intact core samples. The crystallization damage was notably higher for the tighter rocks. The shear modulus (rigidity) was particularly affected. On average, the decrease in shear modulus for BBs mounts to a half and TRs to a third.
- We conceptualized two potential sets of mechanisms at pore and grain level that can be utilized to describe how salt nucleation, precipitation and growth may contribute to stress localization and mechanical damage in porous reservoir sandstones.
- The pronounced decline in rigidity and observed dilation signals the elevated risk of shear failure (when in-situ shear stress exceeds the shear strength of $CO_2$ storage reservoirs) in the near-wellbore regions during $CO_2$ injection in hypersaline aquifers. In such storage sites, there is a higher probability of extensive $CO_2$-induced salt growth in the evaporation front near the well-reservoir contact, where in addition to associated injectivity problems, the risk of mechanical failure exacerbated by resulting excess pressure buildup can result in integrity issues and economic consequences.

These findings may have critical implications for carbon sequestration projects in (hyper)saline aquifers. The cumulative effects of salt crystallization (coupled chemical-mechanical-hydraulic), particularly in quartz-rich sandstones, present noteworthy challenges to the integrity and economic viability of these operations. This study underscores the complexities of subsurface geochemistry-geomechanics. It emphasizes the importance of understanding, predicting, and managing these processes for the long-term integrity of geological units selected for carbon sequestration.

## Acknowledgments

This work is supported by the "solid and salt precipitation kinetics during $CO_2$ injection into reservoir" project, funded by Norway Grants (Norwegian Financial Mechanism 2014-2021) under grant number UMO-2019/34/H/ST10/00564. H. Derluyn acknowledges the support from the European Research Council (ERC) under the European Union's Horizon 2020 research and innovation programme (grant agreement No 850853). The authors thank Ibrahim Omar Khaled for the assistance in x-ray diffraction analysis, Salahalldin Akhavan for support in sample preparation, and Siri Simonsen for help with energy dispersive spectroscopy.

## Appendix

**Table 1.** Chemo-mechanical interactions of $CO_2$ (dry, wet and water/bine-acidified) with sandstone rock type in natural environments and laboratory settings, including natural seepage (NS), batch (B) and flow-through (F) experiments.

| Literature | Sample | Test Type & Duration | Test Conditions (MPa & °C) | Carbonate Content | Carbonate Type | $\varphi_i$ | $\varphi_f$ | $k_i$ | $k_f$ | $E_i$ | $E_f$ | $K_i$ | $K_f$ | $\nu_i$ | $\nu_f$ | $UCS_i$ | $UCS_f$ | Remarks |
|---|---|---|---|---|---|---|---|---|---|---|---|---|---|---|---|---|---|---|
| (Espinoza et al., 2018) | Summerville (USA) | NS | -- | 23-43 | sparry calcite and micritic calcite cement | 4.8 ± 1.6 | 10.2 ± 0.4 | | | 32,80 | 7,60 | | | 0,22 | 0,15 | 159 | 20,4 | CO2-charged brine induces time-dependent reductions in stiffness, strength, and brittleness in sedimentary rocks. Localized leakage alters deformational behavior and mechanical properties substantially. |
| (Espinoza et al., 2018) | Entrada (USA) | NS | -- | 14 | 5.5: sparry calcite | 8.3±0.4 | 11.6±0.3 | | | 20 | 14 | | | 0,32 | 0,3 | 66,1 | 57 | same as above |
| (Hangx et al., 2015) | Röt Fringe (Netherlands) | NS | -- | 2-29(46) | 0-0.4: calcite, 0.5-9: siderite, and mostly dolomite | 7 | 7 | 0,276 | 0,276 | 19 | 21 | 20,1 | 22,7 | | | 117 | | Pre-CO2 injection rock parameters predict reservoir long-term mechanical behavior, unless CO2 reactions dissolve framework-supporting elements. Concentrated early diagenetic cements, if reactive, may weaken porous sandstones, impacting storage site integrity. Screening for such cements is advised. |
| (Skurtveit et al., 2021) | Entrada (USA) | NS | -- | ? | | 7-15 | 15-25 | 3-200 | 20-3000 | 3-14 | | | | | | 34-65 | 19.7 | Bleaching and Fe-oxide removal show unclear strength reduction indications. |
| (Adam et al., 2015) | Taranki (New Zealand) | B (48 h) | 2.76 & ? | 0.3-32 | | 4.5-11 | | | | | | | | | | | | Matrix weakening impact rivals CO2-water substitution. Pronounced velocity changes occur in larger-grained, high-porosity sandstones, primarily due to dissolution. |
| (Rathnaweera et al., 2017) | Hawkesbury (Australia) | B (12 months) | 4-10 & 35 | 7 & 30 | calcite | 29 & 28 | | 103 & 94 | | | | | | | | 54–57 & 48–52 | | CO2 injection induces strength gain in silica-cemented sandstone aquifers via quartz precipitation. Conversely, carbonate-cemented sandstone formations experience mechanical weakening due to calcite dissolution at grain-to-grain contacts during CO2 interaction. |
| (Marbler et al., 2013) | - Birkigt Triassic sandstone(silicate)<br>- Uder Triassic sandstone(carbonate)<br>- Bebertal Permian sandstone(silicate-carbonate) (Germany) | B (12–35 days) | 10 & 100 | 0<br>25<br>4-5 | calcite | 23–28<br>27<br>12–23 | | | | 10.8–13.1<br>11.9–15.5<br>16.7–22.4 | | | | | | | | Autoclave treatment induces geochemical alterations, particularly in carbonate and sheet silicate cements, impacting the granular structure of sandstones. Exposure to pure scCO2 results in reduced strength, modified elastic deformation, and altered effective porosity compared to untreated sandstones. |
| (Zhao et al., 2015) | Xinghe (China) | B/F (5 days) | 15-25 & 50/100 | 20,7 | 8,6: calcite 12,1: dolomite | 8,4 | 8,2 | 756 | 363 | | | | | | | | | CO2-brine-rock interaction worsens rock permeability and raises displacement pressure. Reduced permeability is linked to altered pore structure, attributed to corrosion, secondary mineral formation, and precipitation. The distribution shift towards medium-sized throats is a key outcome. |
| (An et al., 2022) | Laizhou (China) | B/F (1-7 days) | 15 & 200 | 20,1 | calcite | 13,1 | | | | 7,1 | 4-5.5 | | | | | 63,7 | 27-48 | SC-CO2 alters rock strength, decreasing quartz, increasing feldspar. Calcite dissolution dominates in 7 days. Dynamic alteration weakens soluble mineral dissolution, promoting insoluble mineral spallation. Insoluble minerals control altered rock strength; ISR index mirrors changes; highest reaction order influences damage rate. |
| (Hu et al., 2017) | Zigong (China) | B (6 h) | 10 & 50 | 10.6 | calcite | 8,8 | | 0,042 | | | | | | | | | | Pore-fluid and mineral interactions involve complex ion exchange, adsorption, and chemical dissolution. In CO2-bearing sandstone, brittle failure prevails, while H2O-bearing sandstone exhibits plasticity. |
| (Choens et al., 2020) | Boise (USA) | B (1-7 days) | 13,8 & 70 | 9 | calcite | 29 | | | | ~8 | ~8 | | | 0,1 | 0,12 | | | Exposure to hydrous supercritical CO2 weakens without a direct correlation between dissolution and weakening. Quartz-cemented sandstones in storage |
| | | | | | | | | | | | | | | | | | | reservoirs may not see significant changes. Sandstones with vulnerable cementing phases could be adversely affected by hydrous supercritical CO2. |
| (Samuelson and Spiers, 2012) | Hardegsen Bunter (Netherlands) | F | 15 & 113-117 | 8 | dolomite | | | | | | | | | | | | | Supercritical CO2 shows no discernible impact on the coefficient of friction. Short-term effects do not enhance fault zone reactivation, as there's no apparent reduction in frictional strength during CO2 injection-induced stress changes. |
| (Rathnaweera et al., 2015) | Hawkesbury (Australia) | B (4 months) | 8 & 32 | 7 | 5: calcite<br>2: siderite | 37 | | | | 10.3 | 4.2 | | | 0.26 | 0.32 | | | Supercritical CO2 injection induces reservoir rock weakening, attributed to CO2-water-rock mineral interaction, producing carbonic acid and causing quartz corrosion. Additionally, CO2-rock mineral interaction contributes to mineral dissolution in the rock mass. |
| (Foroutan and Ghazanfari, 2020) | Pecos (USA) | F (72 h) | 7.5-21.5 & ? | 7 | calcite | | | | | 23-27 | 21-22 | 6,5 | 5 | 0,17-0.19 | 0,24-0.27 | | | Enriched brine injection induces significant deterioration in elastic properties, heightened plastic deformation, and notable hysteresis behavior. |
| (Foroutan et al., 2021a, 2021b) | Pecos(USA) | F (63 h) | 7.5-21.5 & 25 | 7 | calcite | 4 | 7 | | | 27,5 | 21,7 | 26,2 | 16,6 | | | | | Notable degradation in elastic properties, particularly Young's modulus, is evident, attributable to weakened mineral grain bonding. |
| (Perera et al., 2016) | Quartz-cemented Hawkesbury (Australia) | B (2 years) | 10 & 35 | 6 | 5: calcite<br>1: siderite | 27 | 29 | 90 | 98,1 | 9,6 | 8,3 | | | 0,26 | 0,36 | 26 | 20 | Excessive overburden pressure in saline aquifers during CO2 sequestration may weaken the reservoir rock mass, leading to reservoir subsidence. |
| (Hashemi et al., 2022) | West Delta (USA) | B (24 h) | 10-40 & 42 | 4,5 | ? | 23,1 | | 3,81 | | | | | | | | | | Elastic modulus increased after almost 20 days of scCO2 exposure during injection, possibly due to scCO2 absorption into clay minerals and specimen dehydration. |
| (Zhou et al., 2016) | Red sandstone (China) | F (250 days) | 3-8 & 40 | 4 | calcite | 21 | | 3,1 | 30,1 | 7,9 | 4,9 | 6,9 | | 0,21 | 0,21 | | | Fluid-rock reaction induced notable creep deformation, decreased permeability in CO2-brine creep tests, and led to elastic modulus degradation in indentation tests. |
| (Guen et al., 2007) | Triassic arkosic sandstone (France) | F (70 days) | 8,3 & 40 | 3 (?) | ? | | 15,8 | 17,5 | 461 | | | | | | | | | Percolation of CO2-rich fluids impacts aquifer compaction behavior, influencing long-term CO2 storage and sequestration capacity. |
| (Mikhaltsevitch et al., 2014) | Donnybrook (Australia ) | F (48 h) | 10 & 42 | 2 | siderite | 11.5 | | 0,276 | | | | 16,3 | 12,9 | | | | | P-velocities decrease post-scCO2 injection, aligning closely with the difference between water-saturated and dry samples. Extensional attenuation remains practically unchanged in water-saturated sandstone before and after scCO2 injection. |
| (Raza et al., 2016) | Berea (USA) | F (24 h) | 14 & 35 | 2 | ankerite | 19 | | 420 | | 25,3 | | | | 0,25 | 0,25 | 53 | | Shear velocity and modulus reductions are attributed to clay corrosion and calcite dissolution in the matrix. |
| (Shi et al., 2019) | Mt. Simon (USA) | B/F (7-14 days) | 14-17 & 45-50 | 2 | ankerite and dolomite | 21 | 23 | 9,67 | 34,3 | 14 | 13,2 | 5,2 | 4,9 | | | | | Mt. Simon Sandstone undergoes significant changes in both transport and mechanical properties when exposed to CO2/brine within a short 1–2 week period. |
| (Harbert et al., 2020) | Mt. Simon (USA) | B (30 days) | 13 & 53 | 2 | ankerite and dolomite | 17 | 18 | 2,6 | 8 | 17,1 | 12 | 7,2 | 5,3 | 0,11 | 0,13 | | | CO2/brine-induced geochemical alterations and mineral framework interactions weaken mineral strength at grain boundaries, fostering fracture propagation or generating new fractures along bedding planes. |
| (Harbert et al., 2020) | Mt. Simon (USA) | B (30 days) | 13 & 53 | 2 | ankerite and dolomite | 24 | 21 | 380 | 405 | 16,2 | 19,4 | 7,5 | 10,1 | 0,13 | 0,17 | | | Slight brittleness reduction and -9% porosity change post-CO2/brine exposure. Reduced pore space may contribute to increased local strain. |
| (Tarokh et al., 2020) | Berea (USA) | F (22 days) | 6.9 & 22 | 1 | Fe-dolomite | 25,1 | | 148 | 296 | | | | | | | | | CO2 treatment significantly alters Berea sandstone's poroviscoelastic behavior. Water-saturated specimens exhibit over twice the time-dependent deformation rate under constant load compared to pristine and damaged states. |
| (Sun et al., 2021) | Mt. Simon (USA) | B (21 days) | | 0,8 | calcite | 16 | 22,2 | 9 | 30,3 | 15 | 12,5 | | | | | | | Core incubation in CO2/brine induces mechanical changes, with estimated Young's modulus suggesting potential structure weakening post-exposure. |
| (Hangx et al., 2013) | Captain D (England) | F (1-2 days) | 14 & 20-60 | 0,3 | calcite | 29 | | | | 21,7 | 16,3 | 13,6 | 8,8 | 0,24 | 0,18 | 80,2 | 74,1 | Calcite dissolution, but grain-to-grain contacts are sufficiently quartz-cemented, preventing significant |
| | | | | | | | | | | | | | | | | | | weakening. No shear failure is expected during $CO_2$ injection, minimizing fines production. However, in shallower, less quartz-cemented sandstones, calcite dissolution may pose a risk. Investigating the short-term effects of $CO_2$ injection on mechanical properties is crucial for such formations. |
| (Canal et al., 2014) | Corvio (Spain) | F (24 days) | 8 & 40 | 0,05 | | 14,5 | | | | 11,4 | | | | 0,38 | | 41,3 | | Interlinked hydro-mechanical processes dominate, with a minor contribution from reactive phenomena. |
| (Nooraiepour et al., 2018) | De Geerdalen (Norway) | F (+100 h) | 9 & 19-21 | minor | | <2.5 | | <0.01 | | 29,37 | | | | | | 139,23 | | $CO_2$ injection-induced drying in swelled sandstone reinstated initial fracture permeability. |
| (Rimmele et al., 2010) | Adamswiller (France) | F (30 days) | 28 & 90 | 0 | | 23 | 26 | 10,8 | 108 | | | 6 | 6 | 0,25 | 0,25 | 25 | 25 | Mechanical properties remain unchanged, while porosity increases after $CO_2$ exposure. |
| (Vanorio et al., 2011) | Fontainebleau (France) | F (?) | 15 & 25 | 0 | | 15 | 12 | 1720 | 1380 | | | | | | | | | $CO_2$ injection induces chemo-mechanical alterations in reservoir rocks, impacting baseline transport and elastic properties. Consideration of these changes is essential for accurate interpretation of 4D seismic anomalies in rock-physics models. |
| (Nover et al., 2013) | Bernburg,Volperiehauser, Neidenbach (Germany) | B (10-180 days) | 10-20 & 100/200 | 0 | | 15 | 16 | 1,1 | 3,3 | | | | | 0,26 | 0,36 | | | Low-frequency electrical conductivity experiments show a notable increase, attributed to dissolution at narrow pore throats enhancing pore system interconnection. A phase angle shift suggests alterations in the geometry of the pore-surface area. |
| (Falcon-Suarez et al., 2016) | Synthetic | F (172 h) | 8.2 & 35 | 0 | | 26 | | 1 | 1,32 | | | | | | | | | Velocity changes and energy loss observed with $CO_2$ saturation are linked to heterogeneous fluid distribution within the sandstone pores, not internal solid skeleton discontinuities. |
| (Rinehart et al., 2016) | Tuscaloosa (USA) | F (<1 day) | 13 & 100 | 0 | | 19 | 19 | | | | | | | | | 65 | 35 | Creep strain rates accelerate in facies with more chlorite cements in brine-$CO_2$ solutions. Weakened grain-to-grain contacts lead to earlier-than-expected initial yield, associated with chemically-controlled failure envelope lowering under acidic conditions induced by $CO_2$ in pore solutions. |
| (Tariq et al., 2018) | Berea (USA) | B (10-120 days) | 8.27 & 120 | 0 | | 17 | 17 | 199 | 199 | 32,08 | 20,16 | | | 0,1 | 0,1 | | | $CO_2$ significantly affected petrophysical and mechanical parameters of the studied rocks. |
| (Huang et al., 2020) | Zunyi (China) | F (24 h) | 10 & 32 | 0 | | 18 | | 900 | | | | | | | | 40 | 30 | Brine-sc$CO_2$ co-saturation lowered UCS, Brazilian tensile strength, and fracture toughness in sandstone compared to brine-saturated specimens. sc$CO_2$ had no significant impact on fracture behavior. Strength reduction was attributed to altered pore structure from clay-cementation (kaolinite) dissolution in acidic solution. |
| (Kim and Makhnenko, 2021, 2020) | Berea (USA) | B&F (21 days) | 7 & 22 | 0 | | 22 | 22 | 100 | | | | 30 | 30 | | | | | $CO_2$-water mixture has negligible impact on silica-rich rock and quartz compressibility. However, for calcite-rich rocks and calcite, bulk moduli decrease by 15–21%. |
| (Sun et al., 2020, 2019) | Datong (China) | B (10-30 days) | 8 & 40 | 0 | | | | | | | | | | | | 64.7 | | Sc$CO_2$ + water immersion time reduces sandstone fracture toughness. Prolonged saturation induces pore and crack formation, progressively deteriorating the structure and weakening fracture resistance. |
| (Liu et al., 2014) | Zigong (China) | B | 1-6 & 25 | 0 | | | | 86 | | | | | | | | | | Tensile fracturing of sandstone remained largely unaffected by gaseous $CO_2$, with or without the combined effect of water. |
| (Schimmel et al., 2022) | Bentheim (Germany) | B | 10 & 80 | 0 | | 21.8–23.3 | | | | | | | | | | | | Injection of acidic fluids is likely to impede reservoir compaction. |
| (Akono et al., 2020) | Mt. Simon (USA) | F/B (5/7 days) | 8,6/17,2 & 50-53 | 0 | 0 | 23-40 | | | | | | | | | | | | $CO_2$-induced geochemical reactions lead to microstructural changes, causing a significant decrease in the macroscopic logarithmic creep modulus. |
| (Zhang et al., 2023) | Sichuan (China) | B (7 days) | 0.1 & 25 | 0 | | 6 | 7.5 | | | 9.5 | 8 | | | | | 77 | 65 | Acidic solution effectively reduced compressive strength and elastic modulus in the specimens. |
| (Erickson et al., 2015) | Birkigt Triassic sandstone(silicate) Uder | B (?) | 10 & 100 | ? | ? | 5-25 | | | | | | | | | | 36,1 | 40,4 | Visible or measurable changes in mineral surfaces and pore fluid composition result from rock interaction |
| | Triassic sandstone(carbonate) Bebertal Permian sandstone(silicate-carbonate) (Germany) | | | | | | | | | | | | | | | | | with scCO2, with or without impurities. These compositional changes notably affect rock deformability and compressive strength. In most cases, impurities decrease deformation modules and the maximum capable stress difference. |
| (Zheng et al., 2015) | ? | B | 1 & 25 | ? | ? | | | | | | | | | | | | | CO2 addition lowered shear dilatancy threshold stress by approximately 17%. SEM examination post seepage-creep testing revealed sandstone volumetric deformation due to internal micro-crack extension, particle occlusion, and relative sliding between particles. |
| (Wu et al., 2018) | Pennsylvanian Morrow B (USA) | F (+35 days) | 29 & 71 | ? | ankerite-siderite-cemented / calcite-cemented | | ~8-21% decline | | | | | | | | | | | In all samples, there's a mechanical degradation (E: -31-21%) and a decrease in P- and S-wave velocities. Ankerite-siderite-cemented cores show less degradation compared to control samples. Calcite-cemented cores exhibit significant degradation, especially upstream, relevant to the injection well. Studying a worst-case scenario (100-500 pore volumes injection) for carbon sequestration sites. |
| (Lamy-Chappuis et al., 2016) | Cayton bay (England) | B | 3.4-27.5 & 50 | ? | calcite | 32-34 | 34-38 | | | | | | | | | | | Compelling evidence highlights the indispensability of considering fluid-rock interactions in assessing the mechanical properties of calcite-bearing reservoirs in the context of GCS. |

* φ: porosity (%), k: permeability (mD), E: Young's modulus (GPa), K: bulk modulus (GPa), ν: Poisson's ratio, UCS: uniaxial compressive strength (GPa) where i: initial and f: final.